\documentclass{amsart}

\usepackage{amsfonts}
\usepackage[foot]{amsaddr}
\usepackage{graphicx}
\usepackage{amsmath,amssymb,mathtools,latexsym}
\usepackage{algorithmicx}
\usepackage{algorithm}
\usepackage[noend]{algpseudocode}
\usepackage{algpseudocode}
\usepackage{array}    
\usepackage{url,tikz,pgfplots,xcolor}
\usetikzlibrary{spy,calc}
\usepackage{lipsum,adjustbox}
\usepackage{subcaption}
\usepackage[font={small,singlespacing}]{caption}

\algrenewcommand\alglinenumber[1]{\footnotesize #1}
\algrenewcommand\algorithmicindent{0.9em}%
\algblockdefx{ForAllp}{EndForAllp}[1] {\textbf{in parallel for} #1}%

\makeatletter
\ifthenelse{\equal{\ALG@noend}{t}}%
  {\algtext*{EndForAllp}}
  {}%
\makeatother

\makeatletter
\renewcommand{\email}[2][]{%
  \ifx\emails\@empty\relax\else{\g@addto@macro\emails{,\space}}\fi%
  \@ifnotempty{#1}{\g@addto@macro\emails{\textrm{(#1)}\space}}%
  \g@addto@macro\emails{#2}%
}
\makeatother

\newcommand{\supersuperscript}{
    \hspace{-4.5pt}
    \begin{tikzpicture}
        \draw (0,0) (0,8pt) node {\scriptsize $T\mkern-6mu$};
    \end{tikzpicture}
    }

\DeclarePairedDelimiter\norm{\lVert}{\rVert}%
\newcommand{\batch}[1]{{#1_{\scriptscriptstyle \vert\kern-0.24ex\vert\kern-0.24ex\vert}}}

\title{Hierarchical Matrix Operations on GPUs: Matrix-Vector Multiplication and Compression}

\author{Wajih Halim Boukaram$^1$}
\author{George Turkiyyah$^2$}
\author{David E. Keyes$^1$}

\email{wajihhalim.boukaram@kaust.edu.sa, gt02@aub.edu.lb, david.keyes@kaust.edu.sa}

\address{$^1$Extreme Computing Research Center (ECRC), King Abdullah University of Science and Technology (KAUST), Thuwal 23955, Saudi Arabia.}
\address{$^2$Department of Computer Science, American University of Beirut (AUB), Beirut, Lebanon.}

\begin{document}

\pagestyle{empty}

\begin{abstract}
Hierarchical matrices are space and time efficient representations of dense matrices that exploit the low rank structure of matrix blocks at different levels of granularity. The hierarchically low rank block partitioning produces representations that can be stored and operated on in near-linear complexity instead of the usual polynomial complexity of dense matrices. 

In this paper, we present high performance implementations of matrix vector multiplication and compression operations for the $\mathcal{H}^2$ variant of hierarchical matrices on GPUs. The $\mathcal{H}^2$ variant exploits, in addition to the hierarchical block partitioning, hierarchical bases for the block representations and results in a scheme that requires only $\mathcal{O}(n)$ storage and $\mathcal{O}(n)$ complexity for the mat-vec and compression kernels. These two operations are at the core of algebraic operations for hierarchical matrices, the mat-vec being a ubiquitous operation in numerical algorithms while compression/recompression represents a key building block for other algebraic operations, which require periodic recompression during execution.

The difficulties in developing efficient GPU algorithms come primarily from the irregular tree data structures that underlie the hierarchical representations, and the key to performance is to recast the computations on flattened trees in ways that allow batched linear algebra operations to be performed. This requires marshaling the irregularly laid out data in a way that allows them to be used by the batched routines. Marshaling operations only involve pointer arithmetic with no data movement and as a result have minimal overhead.

Our numerical results on covariance matrices from 2D and 3D problems from spatial statistics show the high efficiency our routines achieve---over $550$ GB/s for the bandwidth-limited matrix-vector operation and over $850$ GFLOPS/s in sustained performance for the compression operation on the P100 Pascal GPU. 

\end{abstract}

\maketitle
\vspace*{2em}

\section{Introduction}
Large dense matrices are ubiquitous in scientific computing. The discretization of integral operators associated with elliptic PDEs results in systems that are dense and on the order of the mesh size. Schur complement methods exploiting dimension reduction in PDE discretizations give rise to large dense systems. Kernel-based machine learning algorithms generate large dense matrices describing pairwise relations between data points. Numerical optimization problems arising in inverse problems and data assimilation are generating ever-more exigent demands for manipulating large dense Hessians. Spatial statistics generates dense covariance matrices from ever larger data sets.

The sizes of these matrices as they arise in practical applications make their direct storage prohibitive and would require algorithms of polynomial complexity for performing matrix-vector multiplication, matrix-matrix multiplication, factorization, and related linear algebra operations. Fortunately, many of these matrices described above have an underlying \emph{data sparse} structure, consisting of blocks many of which can be well-approximated by low rank factorizations. Even though the blocks are of varying sizes and locations in the matrix, tree-based data structures can be used to take advantage of this inherent data sparsity and organize the block approximations hierarchically, in effect compressing the dense matrix in an accuracy-controlled manner. The resulting representations, termed hierarchical matrices, provide an efficient and practical way of storing the dense matrices of very large dimension that appear in a broad range of settings. 

Hierarchical matrices can avoid superlinear growth in memory requirements and store $n\times n$ dense matrices in a scalable manner. For the $\mathcal{H}^2$ hierarchical representations considered in this paper, they require only $\mathcal{O}(kn)$ units of storage where $k$ is a representative rank for the low rank blocks. This asymptotically optimal storage requirement of hierarchical matrices is a critical advantage, particularly in GPU environments characterized by relatively small global memories. For many standard applications, the compressed hierarchical form produces a few orders-of-magnitude reduction in required memory compared to the equivalent dense representation and makes it possible to fit the matrix in the limited global memory of current generation GPUs, overcoming the disadvantage of the slow transfer of data between GPU and main memory. 

Efficient CPU hosted algorithms and software for hierarchical matrices are available~\cite{hlibpro} and have been used in a variety of applications. More recently a task-based parallel implementation was demonstrated on the Intel Phi~\cite{kriemann13}. In contrast, there have been only limited efforts in the development of algorithms appropriate for GPU environments. For example, a recent work accelerated some of the readily vectorizable portions of the computation, such as setting up an initial stiffness matrix \cite{borm15}. Another work used parallel work queues for $\mathcal{H}$-matrix vector multiplication \cite{pzaspel17}. However, methods addressing the core $\mathcal{H}^2$-matrix operations on GPUs are not yet available.  

The lack of high-performance GPU algorithms is likely due to the fact that the naturally recursive data structures and formulations of the hierarchical matrix algorithms do not readily map to the throughput-oriented architecture of GPUs. Alternative representations and algorithmic formulations are needed to exploit the memory hierarchy of GPUs, expose fine-grained parallelism, orchestrate data movement to hide latencies and reduce global memory transactions, and increase occupancy to enhance parallelism, in order to obtain performance. Because hierarchical matrices occupy conceptually a middle ground between dense and sparse matrices, they can inherit some of the powerful GPU advantages of working with regular memory access patterns and can also leverage ideas from algorithms for sparse linear algebra computations on GPUs~\cite{filippone17,bell12,bell09,merrill16} for working efficiently with the irregular patterns. 

This work seeks to develop GPU-resident data structures and associated data parallel algorithms for operating on hierarchical matrices.  Specifically, we describe two algorithms including matrix-vector multiplication (HMV) and matrix compression that operate on flattened representations of the hierarchical matrix.  
Both algorithms are work optimal, $\mathcal{O}(n)$, and demonstrate high absolute performance on matrices of size up to $1M\times1M$ stored entirely on the GPU. The memory-bound HMV achieves more than $550\,$GB/s, surpassing the STREAM benchmark \cite{stream}, and the compute-bound hierarchical matrix compression achieves more than $850\,$GFLOPS/s on the Pascal P100 GPU. These two operations are foundational routines for almost all other algebraic operations on hierarchical matrices, including matrix multiplication, inversion, factorization and others. We plan to use them as the building blocks for a complete GPU $\mathcal{H}^2$-library. We also hope that by making available high performance implementations of these basic hierarchical matrix routines, we will encourage broader experimentation with hierarchical matrices in various applications. We employ the word ``experimentation'' advisedly, inasmuch as the numerical analysis of rank growth and error propagation in chains of hierarchical operations is not yet completely mature. It may be that the high compressibility will prove more practically tolerable in some applications than others.

The rest of this paper is organized as follows. In Section 2, we describe the flattened data structures used to represent the row and column basis trees as well as the matrix tree that stores the matrix block data expressed in terms of these row and column bases. Section 3 describes a GPU matrix vector multiplication (HMV) algorithm and shows its performance on sample covariance matrices arising from 2D and 3D spatial statistics. Section 4 describes the hierarchical compression operation expressed in terms of batched QR and SVD operations, and analyzes the performance of its various phases on the same covariance matrices above. Discussion and conclusions are presented in Section 5. 


\section{Hierarchical matrices}

\subsection{Flavors of hierarchical matrices} 

A great deal of work has been done in the development of representations that exploit the low rank structure of matrix blocks in a hierarchical fashion. We do not attempt to review this literature here except for mentioning a few representative works in this section. We refer the reader to \cite{hackbusch15,ballani16} for an introduction and survey. 

Hackbusch \cite{hackbusch99,hackbusch99} pioneered the concepts of hierarchical matrices in the form of $\mathcal{H}$ and $\mathcal{H}^2$ matrices as a way to generalize fast multipole methods, and developed a substantial mathematical theory for their ability to approximate integral operators and boundary value problems of elliptic PDEs. These ideas have been developed considerably over the years, for the construction and use of hierarchical matrices in solving discretized integral equations and preconditioning finite element discretizations of PDEs ~\cite{borm02,borm05,borm2010,grasedyck2003,grasedyck2009}.

Hierarchically semi-separable (HSS) and hierarchically block-separable (HBS) are related and well-studied rank-structured representations that also use low rank blocks of a dense matrix in a hierarchical fashion. 
Matrices are semi-separable if their upper and lower triangular parts are, each, part of a low rank matrix. HSS matrices extend this idea and refer to matrices whose off-diagonal blocks are all of low rank and expressed in a nested basis. Their block structure is equivalent to what is also known as a weak admissibility criterion~\cite{hackbusch04}. HSS matrices have been shown to be useful representations for integral equations in the plane and for sparse systems of equations that can be reduced to matrices of this form, for example by using nested dissection on 2D grids. Fast factorization algorithms for HSS matrices have been developed in~\cite{xia10}. HBS matrices have a similar structure but emphasize the telescoping nature of the matrix factorization \cite{gillman2012direct} to use in the construction of direct solvers for integral equations \cite{Martinsson2013}. Hierarchically off-diagonal low rank (HODLR) matrices which simplify the HSS representation by using non-nested and separate bases for various matrix blocks have been used for fast factorizations in \cite{ambikasaran2013n}.


\begin{figure}[t]
  \begin{center}
	\includegraphics[width=0.5\textwidth]{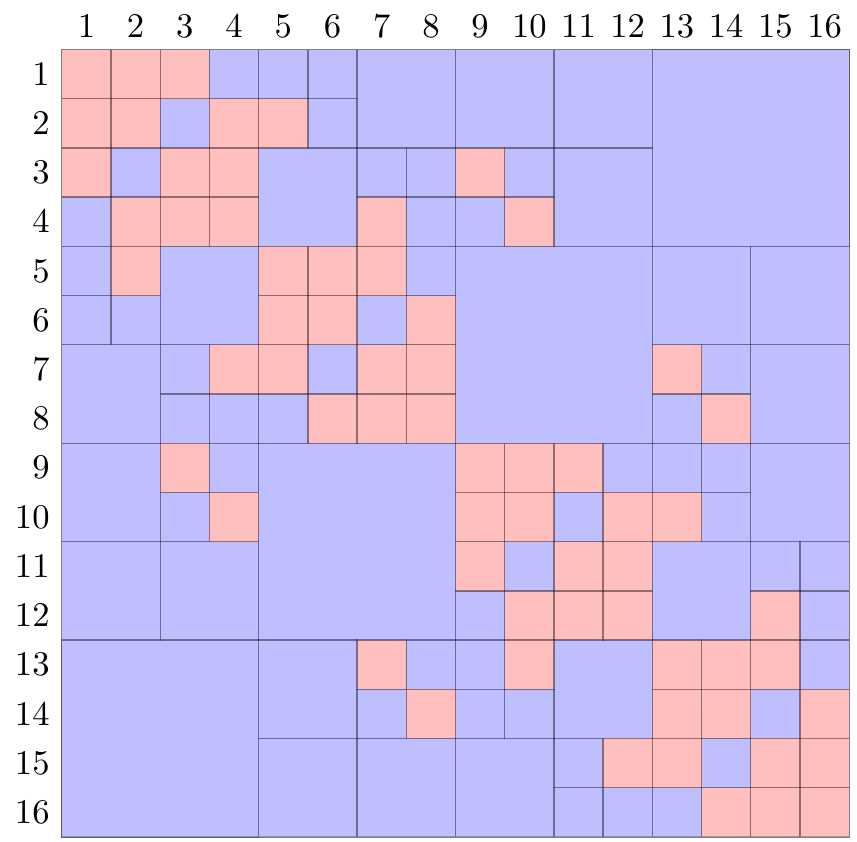}
	\caption{A three-level hierarchical matrix with its dense $m \times m$ blocks shown in red and its low rank blocks shown in blue.}
	\label{fig:h2mat}
  \end{center}
\vspace*{1em}
  \includegraphics[width=\textwidth]{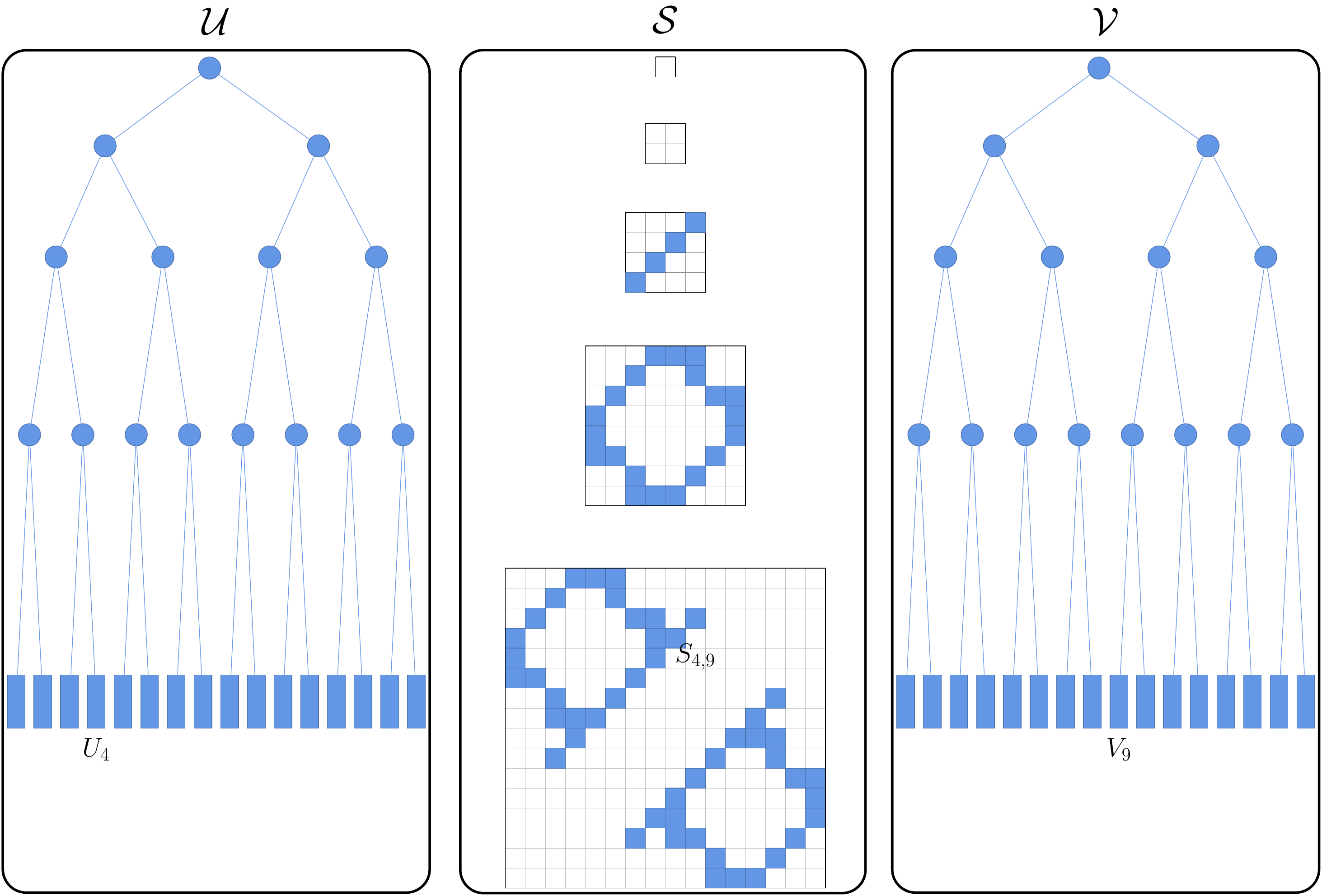}
  \caption{The low rank part of the matrix in Fig.~\ref{fig:h2mat} ``disassembled'' into its constituent basis trees ($\mathcal{U}$ and $\mathcal{V}$) and matrix tree ($\mathcal{S}$) representations. Representation is done level by level for all trees.}
  \label{fig:h2tree}
\end{figure}

\subsection{Structure of a general hierarchal matrix with nested bases}
\label{sec:hmatrix_structure}
In this paper, we use the $\mathcal{H}^2$ representation as it allows a general block structuring of the matrix, has asymptotically optimal memory complexity, and in practice results in a representation with a small memory footprint in applications.  The representation achieves the $\mathcal{O}(n)$ memory complexity by exploiting two different types of hierarchies in the underlying matrix. 

One type of hierarchy is related to the granularity of matrix blocks where larger blocks that admit low rank approximations sit at higher levels in a tree representation of the matrix, whereas the smaller low rank blocks sit at the lower levels. Figure 1 illustrates a block partitioning of a sample matrix. The matrix has three different block sizes that admit a low rank representation and are shown in blue. The matrix also has some blocks that do not admit such a representation and are stored as dense matrices and are shown in red. We denote the low rank decomposition of a given block by $U S V^T$, where $S$ is a $k\times k$ matrix with $k$ generically denoting the rank of the block, and refer to $U$ and $V$ as the column and row bases in which the $S$ block data is expressed. The $S$ matrices are termed coupling matrices. 

The middle diagram of Figure 2 shows the levels of the matrix separated out. The bottom level contains the $S$ data of only the smallest blocks of the matrix, the next level up contains the $S$ data corresponding to the mid-sized blocks, and the level above it contains the $S$ data for the largest blocks of the partitioning. The top two levels of the tree illustrated here are empty because there are no blocks of the appropriate size  that admit a low rank representation. This hierarchy of block partitioning is common to all hierarchical matrix formats, although the $\mathcal{H}$ and $\mathcal{H}^2$ representations offer the most flexibility and generality, as they do not place restrictions on the admissible partitionings. Low rank blocks of any size, as well as dense blocks, can be located anywhere in the matrix. 


The second hierarchy, specific to the $\mathcal{H}^2$ format, is related to the manner in which the low rank  blocks are represented, using nested row and column bases, $U$ and $V$. Nestedness means that bases for the higher levels (larger blocks) may be obtained through suitable transformations of the bases of the lower levels and therefore need not be explicitly stored. Figure 2 illustrates how the bottom level of the basis trees is explicitly stored and can be directly used. For example, the block $(4,9)$ of the matrix of Figure 1 is a low rank block of the smallest size, therefore expressed as $U_4 S_{49} V_9^T$, with $U_4$ and $V_9$ explicitly stored at the leaves of the bases trees as shown. At the next level up, the bases do not have an explicit representation but each basis (denoted graphically by a circle) can be obtained, when needed, from its children through small transfer matrices. These nested bases allow significant reduction in storage and produce the algorithmically optimal memory complexity. 

%

The representation of the low rank portion of a hierarchical matrix $A_{n\times n}$ consists therefore of the tree triplet $\mathcal{U}$, $\mathcal{S}$, $\mathcal{V}$:
\begin{itemize} 
	\item The $\mathcal{U}$ tree organizes the row indices of the matrix hierarchically. We use a binary tree in this work but other organizations are possible. A node in the tree at level $l$ represents a row block at this level of the matrix and is used to store column basis vectors in which the data of the matrix blocks at level $l$ are expressed. Thin basis matrices $U$ of size $m \times k$ are stored explicitly at its leaves. Small interlevel transfer matrices $E$ of size $k_c \times k_p$ (referring to the ranks of the child and parent nodes) are stored at the higher levels and used to compute with the level-appropriate bases, which are never explicitly stored. When referring to a basis node as $U_i^l$, we refer to the node $i$ at level $l$ in the basis tree which is either stored explicitly at the leaves, or implicitly via the transfer matrices higher level of the tree. The relationship between a node $U_{i^+}^{l-1}$ and its children $U_{i_1}^l$ and $U_{i_2}^l$ uses the transfer matrices $E_{i_1}^l$ and $E_{i_2}^l$: 
\begin{equation} 
\label{eq:transfer}
	U_{i^+}^{l-1} = 
	\begin{bmatrix} U_{i_1}^l & \\ & U_{i_2}^l \end{bmatrix}
	\begin{bmatrix} E_{i_1}^l \\ E_{i_2}^l \end{bmatrix}. 
\end{equation} 
Similarly, the $\mathcal{V}$ tree, consisting of explicit thin basis matrices at the leaves and small inter-level transfer matrices $F$, organizes column indices hierarchically and its nodes are used to represent the row basis vectors for the column blocks at various level of granularity. The structure of this tree need not be identical to that of $\mathcal{U}$, although the examples described in this paper come from symmetric matrices where we use the same block row and column trees.  

	
	\item The $\mathcal{S}$ tree is an incomplete quadtree that stores the coupling matrices $S$ for all the blocks of the matrix. Other N-ary trees would be needed if the basis trees were not binary. As we describe below, the tree is stored level by level, with each level being a block sparse matrix. The sparsity pattern of the block sparse matrix at level $l$ is directly related to the low ranks blocks that exist at that level of granularity: a $k\times k$ coupling matrix exists in the entry $(i,j)$ of level $l$ if a block at that level of granularity exists in the hierarchical matrix partitioning. In that case, the corresponding matrix block is $U_i^l S_{ij}^l {V_j^l}^T$, where $U_i^l$ and $V_j^l$ are the column bases and row bases of block row $i$ and column block $j$ at level $l$, respectively. The non-zero block entries from all levels form the leaves of the quadtree, and collectively they cover all the low rank blocks of the matrix. The storage needed for $\mathcal{S}$ depends on the structure of the tree and the distribution of the leaves but assuming a bounded number of non-zero entries in the block rows/columns, a reasonable assumption in many applications, its memory requirements have optimal complexity, $\mathcal{O}(kn)$, where $k$ is a representative rank.  
\end{itemize}

Besides the low rank blocks, a set of dense $m \times m$ matrices that are not compressed need to be also stored. The complement of the low rank leaves of the quadtree represents blocks of the original matrix that are not economically expressible as low rank factorizations and are more effectively stored in their explicit dense  format. These dense leaves appear only at the finest level of the quadtree and in practical terms represent blocks of a fixed small size that is tuned to the execution hardware. We have used $m = 64$ for the examples in this paper. We store these dense blocks as a separate block sparse matrix and allow them to appear anywhere in the matrix. 

Notation. Table 1 summarizes the symbols used in the description of the tree algorithms on the hierarchal matrix. 

\begin{table}[!t] 
\footnotesize
{\begin{tabular}{l c l} 
	\hline 
	Symbol & & Description \\ \cline{1-1} \cline{3-3}
	$n$ & & matrix size \\ 
	$m$ & & size of dense blocks \\ 
	$k^{l}$ & & typical rank of blocks in matrix tree at level $l$ \\ 
	$\mathcal{U}, \mathcal{V}$ & & block row and column basis trees, with explicit bases stored at leaves only\\ 
	$U$, $V$ & & bases at the leaf level of $\mathcal{U}$ and $\mathcal{V}$ \\
	$E$, $F$ & & transfer matrices for the  $\mathcal{U}$ and $\mathcal{V}$ bases, respectively \\
	$\mathcal{S}$ & & matrix quadtree of coupling matrix blocks \\ 
	$i$, $j$ (or $i_1, i_2$, etc.) & & indices of block rows and block columns respectively \\
	$i^+$ & & index of the parent block of block $i$ \\
	$x(i)$ & & sub-vector of a vector $x$ corresponding to the block $i$ \\
	$\hat{x}, \hat{y}$ & & vectors defined at every level in the basis trees \\
	$\batch{U}$, $\batch{E}$, etc. & & batched $U$, $E$, etc.~arrays, marshaled for use by batched linear algebra kernels \\
	\hline 
\end{tabular}} 
\caption{Notation used.} 
\label{table:notation} 
\end{table}

\section{Strategies for Efficient GPU processing of $\boldsymbol{\mathcal{H}}$-matrix Trees}
\label{sec:hmatrix_flattening}
GPU routines are executed as kernels and can be called from the host CPU by specifying a launch configuration, organizing the GPU threads into thread blocks and thread blocks into grids. Launching a kernel causes a short stall (as much as 10 microseconds) as the kernel is prepared for execution. Let us call this stall the kernel launch overhead. For kernels that do a lot of processing, this stall is quite insignificant and does not impact performance; however, when the kernel execution is comparable to the overhead it presents a problem. All of the individual operations involved in $\mathcal{H}$-matrices are on very small matrix blocks. Execution of these operations using a kernel call per block will be limited by the kernel overhead. To minimize the impact of the overhead, operations are best executed in large batches \cite{batch_haidar}. Efficiently marshaling the operations into these batches is also key to a high performance implementation. To that end, we flatten the tree by levels and store the level data in contiguous chunks of memory where each matrix block is stored consecutively in column major order. Operation marshaling then involves specialized kernels that process each level in parallel to produce the necessary data for the linear algebra batch kernels. The benefits of this decomposition are two-fold: the marshaling kernels can access the level data in coalesced global memory reads and the batch kernels can execute without any knowledge of the tree data structure. Since every kernel call uses a single launch configuration, the operations handled by each batch routine must have the same size; in the $\mathcal{H}$-matrix setting, this translates to a fixed rank per level of the trees. Variable size batch kernels will be needed to overcome this limitation, but this will be the focus of future work. 

\subsection{Flattened Tree Structure}
The flattened structure of the tree is represented by three arrays $head$, $next$, and $parent$ of node indices, with each node index referring to a block of rows/columns. The $head$ array contains the node index of the first child of a node while each entry in the $next$ array gives us the index of the next sibling of a node. The $parent$ array contains the parent node index of each node allowing us to traverse the tree in any direction. An example of this storage scheme for the structure of the basis tree is depicted in Figure \ref{fig:basis_tree_mem}. The node indices stored in this flattened structure are used by the marshaling routines of the various hierarchical operations to efficiently generate data from each level that can then be passed on to the batch kernels. The data can come from trees that share the same structure as the basis trees, such as the $\widehat{x}$ and $\widehat{y}$ trees of the MVM described in Section \ref{sec:hmvm_overview} and the $Z$ and $T$ trees of the orthogonalization described in Section \ref{sec:basis_orthogonalization}.

\begin{figure}[!t]
	\begin{center}
	\begin{adjustbox}{width=0.9\linewidth}
		\begin{tabular} {c}
			\tikzset{
				treenode/.style = {align=center, inner sep=0pt, text centered}, 
				node/.style = {treenode, circle, black, draw=black, text width=1.5em, thick},
				leaf/.style={rectangle,thick,draw,text width=0.35cm, text height = 0.7cm, text centered},
				level 1/.style={sibling distance=40mm},
				level 2/.style={sibling distance=20mm},
				level 3/.style={sibling distance=10mm},
				level 4/.style={sibling distance=5mm}
			}
			
			\begin{tikzpicture}[-,level/.style={level distance = 1.0cm}] 
			\node [node] {1}
			child{ node [node] {2} 
				child{ node [node] {4} 
					child{ node [node] {8} {child {node [leaf] {1}}} }
					child{ node [node] {9} {child {node [leaf] {2}}} }
				}
				child{ node [node] {5}
					child{ node [node] {10} {child {node [leaf] {3}}}}
					child{ node [node] {11} {child {node [leaf] {4}}}}
				}                            
			}
			child{ node [node] {3}
				child{ node [node] {6} 
					child{ node [node] {12} {child {node [leaf] {5}}} }
					child{ node [node] {13} {child {node [leaf] {6}}} }
				}
				child{ node [node] {7}
					child{ node [node] {14} {child {node [leaf] {7}}} }
					child{ node [node] {15} {child {node [leaf] {8}}} }
				}
			}
			; 
			\end{tikzpicture}
			\\
			\\
			\def\arraystretch{1.5}
			\newcolumntype{C}{>{\centering\arraybackslash}p{2.5ex}}
			\begin{tabular}{|c|C|C|C|C|C|C|C|C|C|C|C|C|C|C|C|}
				\hline
				Level Pointers & 1 & 2 & 4 & 8 & 16 &&&&&&&&&& \\ \hline
				$parent$ & -&1&1&2&2&3&3&4&4&5&5&6&6&7&7 \\ \hline
				$head$   & 2&4&6&8&10&12&14&-&-&-&-&-&-&-&- \\ \hline
				$next$   & -&3&-&5&-&7&-&9&-&11&-&13&-&15&- \\ \hline
			\end{tabular}
		\end{tabular}
	\end{adjustbox}
	\end{center}
	\caption{Basis tree structure. 
		Data associated with the basis tree uses the node numbers to locate the position of the node data in memory. 
		Examples for the MVM operation include the $\hat{x}$ and $\hat{y}$ trees (in the nodes), 
		the basis leaves (shown as rectangles), and the inter-level transfer matrices (in the nodes).}
	\label{fig:basis_tree_mem}
\end{figure}

\subsection{Marshaling Operations}
Specialized marshaling kernels for each operation generate the data that batched linear algebra routines need to execute the operation. This includes the pointers to matrix blocks from a level as well as matrix block dimensions and we denote marshaled data by the $|||$ subscript. This data can then fed into the appropriate batch routines for execution using a single kernel call per level. Since the levels of the tree have been flattened into arrays, it is straightforward to parallelize using simple parallel transformations, either by a simple kernel or using libraries such as Thrust~\cite{bell2011thrust}. These transformations are executed very quickly and constitue a negligible portion of the execution time within each operation. Examples of the marshaling routine for a few of these operations are described in each section, such as the upsweep marshaling of the matrix vector operation described in Algorithm \ref{alg:hgemv_marshalupsweep} and the marshaling of the orthogonalization operation described in Algorithm \ref{alg:horthog_marshalupsweep}, with a few omitted for the sake of brevity. For the operations presented in this paper, we rely on high performance batched linear algebra kernels for QR and singular value decompositions. We do not describe these kernels here, as their details may be found in~\cite{BOUKARAM2017}. We also use the high performance matrix-matrix multiplication batched routines from the CUBLAS~\cite{cublas16} and MAGMA~\cite{dongarra14} libraries.


\section{Hierarchical matrix-vector multiplication}
\label{sec:HMVM}

\subsection{Overview}
\label{sec:hmvm_overview}
In this section, we describe the different phases of the HMV algorithm and their efficient GPU implementations.
\begin{figure}[!ht]
	\begin{center}
		\includegraphics[width=0.85\textwidth]{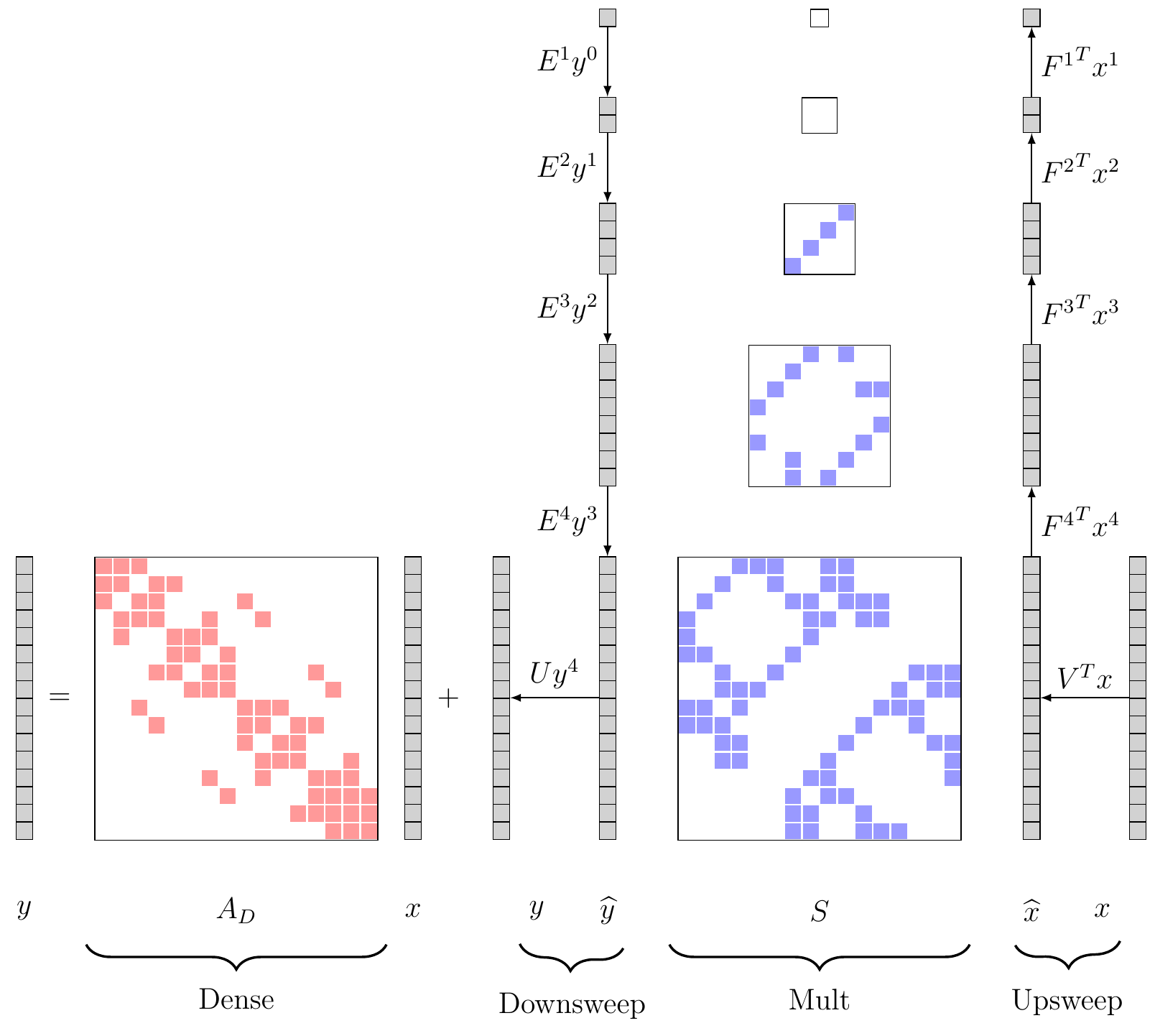}
	\end{center}
	\caption{Overview of the HMV. Computation with the low rank blocks is split into three phases: upsweep, multiplication, and downsweep. The upsweep computes a tree $\hat{x}$ from the input vector $x$ by first projecting it to the basis leaves and then sweeping up the tree using the transfer matrices. $\hat{x}$ is then fed into the block sparse multiplication phase to produce the $\hat{y}$ tree. Finally, the downsweep computes the output vector $y$ by first accumulating partial sums within each level of $\hat{y}$ using the transfer matrices. The leaf level of $\hat{y}$ then contains the full sums in terms of the basis leaves which are expanded to form the output vector $y$ and added to the results of the dense matrix-vector product to produce the final result.}
	\label{fig:hmatvec}
\end{figure}
Since the hierarchical matrix may be viewed as the sum of a dense portion $A_D$ and a low rank portion $A_{LR}$ the products with $A_D$ and $A_{LR}$ are done separately (but concurrently when possible, as detailed below) and added as shown in Figure~\ref{fig:hmatvec}.
The product of the dense blocks with the input vector is computed via a block sparse matrix vector multiplication routine. The product of the low rank blocks with the input is then computed in three additional phases: an upsweep phase, a multiplication phase, and a downsweep phase as illustrated in Figure~\ref{fig:hmatvec}. 
For intuition, we may think of this algorithm as the hierarchical generalization of the way we multiply a regular dense low rank matrix, $USV^T$, by a vector $x$. In this case we would do it in three phases: first apply the transpose of $V$ to $x$, then multiply the small resulting product by $S$, and then apply $U$ to obtain the final product. The hierarchical analogue first applies the transpose of the bases of all levels of $\mathcal{V}$ to $x$ by sweeping up the tree to compute the nodes of a vector tree $\widehat{x}$. 
The multiplication phase then computes a vector tree $\widehat{y}$ where each node in $\widehat{y}^l$ is the product of a block row of coupling matrices in $S^l$ with the corresponding nodes of $\widehat{x}^l$. The nodes of $\widehat{y}^l$ represent the level's partial contribution to the output vector expressed in the row basis $U^l$. Finally, a downsweep phase expands the nodes of $\widehat{y}$, multiplying them by the bases from the corresponding levels of $\mathcal{U}$ to produce the final output vector $y$. It is also worth noting that these phases of the HMV computation are very closely related to the phases of the fast multipole method~\cite{sfi14}.

\subsection{Upsweep phase}
The upsweep computes a tree $\widehat{x}$ as the products of the transposed nodes of the column basis $V$ with the input vector $x$; i.e. $\widehat{x}^{l}_j = {V^{l}_j}^T x(j)$ for all nodes $j$ within a level $l$, with $l$ in $0\cdots q$. This process is trivial for the leaves since they are stored explicitly; however the inner nodes are expressed in terms of their children using the relationship defined in Eq.~\ref{eq:transfer}. For simplicity, let us consider a parent node $j^+$ with two children $j_1$ and $j_2$. The node $\widehat{x}^{l-1}_{j^+}$ can be computed as:
\begin{equation}
\widehat{x}^{l-1}_{j^+} = 
\begin{bmatrix}
F_{j_1}^{l}\supersuperscript  & {F_{j_2}^{l}}\supersuperscript 
\end{bmatrix}
\begin{bmatrix}
{V_{j_1}^{l}}^T &     \\
& {V_{j_2}^{l}}^T \\
\end{bmatrix}
\begin{bmatrix}
x(j_1)\\
x(j_2)
\end{bmatrix} = 
{F_{j_1}^{l}}\supersuperscript  \widehat{x}^{l}_{j_1} + {F_{j_2}^{l}}\supersuperscript  \widehat{x}^{l}_{j_2}.
\end{equation}
We can compute every node in $\widehat{x}$ by starting at the leaves and sweeping up the tree using the above equation. To avoid the prohibitive overhead of $\mathcal{O}(n)$ kernel launches required to execute this operation recursively on the GPU, we use the flattened tree structure described in section \ref{sec:hmatrix_flattening} to compute $\widehat{x}$ level by level. The leaves are processed simply using a single batch matrix vector operation. Considering a binary tree for the basis trees, an upsweep kernel marshals the data for the operations in a level to generate two batches (one for each child) which are then executed by the batch matrix vector operation. Like all marshaling operations used in the rest of this paper, this marshaling operation uses the flattened tree structure described in Section \ref{sec:hmatrix_flattening} to efficiently generate the necessary pointer data for the batched routines, the $\batch{F}$, $\batch{x}$ and  $\batch{y}$ pointer arrays in this operation, using a single kernel call. This leads us to Algorithm \ref{alg:hgemv_marshalupsweep} for marshaling the upsweep operations and Algorithm \ref{alg:upsweep2} to compute $\widehat{x}$.

\begin{algorithm}[t]
\caption{GPU upsweep marshaling routine}
\label{alg:hgemv_marshalupsweep}
\begin{algorithmic}[1]
\Procedure{marshalUpsweep}{$F^{(l)}$, $\widehat{x}^{(l)}$, $\widehat{x}^{(l-1)}$}
	\State $n_p$ = levelptr$[l-1]$
	\State $k_p$ = levelrank$[l-1]$
	\State $n_c$ = levelptr$[l]$
	\State $k_c$ = levelrank$[l]$	
	\ForAllp{$p = n_p \rightarrow n_c$} 
		\State $i = p - n_p$     \Comment{{\small Batch index}}
		\State $c$ = head$[p]$ 
		\State $c_i = 0$
		\While {$c \neq \text{empty}$}
			\State $\batch{F}(c_i)[i] = $ ptr$\left(F^{(l)}\right) + (c - n_c) \times k_c \times k_p$ 	 \Comment{{\small Extract level pointer data}}
			\State $\batch{x}(c_i)[i] = $ ptr$\left(\widehat{x}^{(l)}\right) + (c - n_c) \times k_c$
			\State $\batch{y}(c_i)[i] = $ ptr$\left(\widehat{x}^{(l-1)}\right) + i \times k_p$
			\State $c_i = c_i + 1$
			\State $c = $ next$[c]$
		\EndWhile
	\EndForAllp
\EndProcedure
\end{algorithmic}
\end{algorithm}

\begin{algorithm}[b]
\caption{GPU upsweep algorithm for forming $\widehat{x}$}
\label{alg:upsweep2}
\begin{algorithmic}[1]
\Procedure{upsweep}{$V$, $F$, $x$, $\widehat{x}$}
	\State $q$ = heightof$( \widehat{x} )$     \Comment{{\small \emph{leaf level, log(n/m)}}}
	\State gemvBatched$\left( \frac{n}{m}, V^T, ( {\tt batch} ) \, x , \widehat{x}^{q} \right)$ 
	\For{$l$ = $q \rightarrow {}1$} 	\Comment{{\small \emph{up the $\mathcal{V}$ tree}}}
		\State $N$ = $n / m / 2^{q-l+1}$
		\State $[\batch{F}^T, \batch{x}, \batch{y}]$ = marshalUpsweep$(F^{l}\supersuperscript, \widehat{x}^{l}, \widehat{x}^{l-1})$	
		\vspace{2pt}
		\For{$j$ = $1 \rightarrow 2$} 		\Comment{{\small \emph{binary tree}}}
			\State gemvBatched$(N, \batch{F}(j)^T, \batch{x}(j), \batch{y}(j))$ 
		\EndFor
	\EndFor
\EndProcedure
\end{algorithmic}
\end{algorithm}

\subsection{Multiplication phase}
The second phase of the computation builds a vector tree $\widehat{y}$, where each node $i$ in a level $l$ is the product of the block row $i$ of level $l$ of the coupling matrix tree with the corresponding nodes in $\widehat{x}$. This operation can be expressed as
\begin{equation}
	\widehat{y}_i^{l} = \sum_{j \in \{b_i\}} S_{ij}^l \widehat{x}_j^{l}
\end{equation}
where $b_i$ is the set of column indexes of the matrix blocks in the block row $i$. We could follow the same marshaling approach as the upsweep, but given the potential nonuniform distribution of blocks in different rows as well as the obvious similarity to a block sparse matrix vector multiplication, we opt to generate block sparse row index data for the matrix tree. This data is efficiently generated once during the construction of the hierarchical matrix and stored per level of the matrix tree. Figure \ref{fig:bsr_example} shows an example of the block sparse row index data for a simple matrix. This leads us to Algorithm \ref{alg:mult} for the computation of $\widehat{y}$. 

\begin{figure}
	\begin{center}
			\begin{tabular}{c}
				$A=\begin{bmatrix}
				a & c &   &   & e & g \\
				b & d &   &   & f & h \\
				&   & i & k &   &   \\
				&   & j & l &   &   \\
				&   &   &   & m & o \\
				&   &   &   & n & p \\
				\end{bmatrix}$ \\
				\\
				\def\arraystretch{1.5}
				\begin{tabular}{|c|c|c|c|c|c|c|c|c|c|c|c|c|c|c|c|c|c|}
					\hline
					Values & \emph{a} & \emph{b} & \emph{c} & \emph{d} & \emph{e} & \emph{f} & \emph{g} & \emph{h} & \emph{i} & \emph{j} & \emph{k} & \emph{l} & \emph{m} & \emph{n} & \emph{o} & \emph{p} \\ \hline
					ColIdx & 1 & 3 & 2 & 3 &&&&&&&&&&&& \\ \hline
					RowPtr & 1 & 3 & 4 & 5 &&&&&&&&&&&& \\ \hline
				\end{tabular}
			\end{tabular}
	\end{center}
	\caption{The BSR storage for a $6\times6$ matrix with $2\times2$ blocks.}
	\label{fig:bsr_example}
\end{figure}

\begin{algorithm}[t]
\caption{GPU Matrix Tree Multiplication for $\widehat{y}$}
\label{alg:mult}
\begin{algorithmic}
\Procedure{TreeMultiply}{$S$, $\widehat{x}$, $\widehat{y}$} 
	\State $q$ = heightof$( \widehat{y} )$    
	\ForAllp{ $l$ = $1 \rightarrow q$}
		\State $\widehat{y}^{l}$ = blockSparseMV$( S^{l}, \widehat{x}^{l})$
	\EndForAllp
\EndProcedure
\end{algorithmic}
\end{algorithm}

\subsection{Downsweep phase}
After the multiplication phase, each level of the vector tree $\widehat{y}$ now contains a partial contribution of the output vector $y$ expressed in terms of the nodes of the block row basis $U$ at that level. We can finalize the computation by expanding the nodes of $\widehat{y}^l$ at each level $l$ as:
\begin{equation}
y(i) = y(i) + U_i^{l} \widehat{y}_i^{l}.
\end{equation}
Since we don't have an explicit representation of the inner nodes of the basis tree, we use the nested basis property to express the partial sum of a level with its child level in terms of the basis nodes of the child level. Taking a parent node $i^+$ at level $l-1$ and its two children $i_1$ and $i_2$ at level $l$, we have the partial sum: 
\begin{equation}
U^{l-1}_{i^+} \widehat{y}^{l-1}_{i^+} + \begin{bmatrix}
U^l_{i_1} \widehat{y}^l_{i_1} \\
U^l_{i_2} \widehat{y}^l_{i_2}
\end{bmatrix} = 
\begin{bmatrix}
U^l_{i_1}  & 		\\
& U^l_{i_2} 
\end{bmatrix}\begin{bmatrix}
E^l_{i_1} \widehat{y}^{l-1}_{i^+} \\
E^l_{i_2} \widehat{y}^{l-1}_{i^+}
\end{bmatrix}
+
\begin{bmatrix}
U^l_{i_1} \widehat{y}^l_{i_1} \\
U^l_{i_2} \widehat{y}^l_{i_2}
\end{bmatrix} = 
\begin{bmatrix}
U^l_{i_1}  & 		\\
& U^l_{i_2} 
\end{bmatrix} \begin{bmatrix}
E^l_{i_1} \hat{y}^{l-1}_{i^+} + \widehat{y}^l_{i_1}\\
E^l_{i_2} \hat{y}^{l-1}_{i^+} + \widehat{y}^l_{i_2}
\end{bmatrix}.
\end{equation}
Sweeping down $\widehat{y}$ and setting each node $\widehat{y}_i^{l} = \widehat{y}_i^{l} + E_{i}^{l} \widehat{y}_{i^+}^{l-1}$, the level $l$ at each step now also contains the partial sum of $y$ for all levels above $l$ expressed in the nodes of $U^l$. The leaf level will then contain the complete sum which is finally expanded into $y$. We follow the same approach as in the upsweep, where each level is processed in parallel by first marshaling the operations and then executing using a batch matrix vector product. This leads us to Algorithm \ref{alg:downsweep} for computing $y$. The downsweep marshaling algorithm is structurally very similar to the upsweep marshaling routine described in Algorithm \ref{alg:hgemv_marshalupsweep} and is omitted for brevity.

\begin{algorithm}
\caption{GPU downsweep for computing $y$}
\label{alg:downsweep}
\begin{algorithmic}[1]
\Procedure{downsweep}{$U$, $E$, $\widehat{y}$, $y$}
	\State $q$ = heightof$( \widehat{y} )$     \Comment{{\small \emph{leaf level, log(n/m)}}}

	\For{$l$ = $1 \rightarrow q$} 	\Comment{{\small \emph{down the $\mathcal{U}$ tree}}}
		\State $N$ = $n / m / 2^{q-l}$
		\State $[\batch{E}, \batch{x}, \batch{y}]$ = marshalDownsweep$(E^{l}, \widehat{y}^{l-1}, \widehat{y}^{l})$		
		\State gemvBatched$(N, \batch{E}, \batch{x}, \batch{y})$ 
	\EndFor
	
	\State gemvBatched$\left( \frac{n}{m}, U, \, \widehat{y}^{q}, y \right)$ 
\EndProcedure
\end{algorithmic}
\end{algorithm}

\subsection{Kernel streaming}
\label{sec_hmv_stream}
The upper levels of the tree operations do not provide enough work to saturate the GPU and kernel overhead starts to impact performance for these levels. To overcome this, we can use streams to try to overlap some stages of the computation with the processing of the upper levels. Unfortunately, the scheduler will only launch a new kernel when the resources are available, which typically happens towards the end of the BSR multiplication kernel. It therefore makes most sense to overlap the dense multiplication portion of the computation with the low rank portion.  Many GPUs support a feature called stream priorities that allow execution of a kernel on a low priority stream to be suspended in favor of a kernel on a higher priority stream. By setting the dense phase as the lowest priority and the tree operations as the highest priority, we can effectively hide the overhead and hardware underuse of the tree operations. The performance results below show the effect of this overlap which, as expected, is beneficial for relatively small sized problems. On larger problems, the work at the higher levels of the trees is a very tiny fraction of the overall computation and there is relatively little benefit derived from the overlap. The $\widehat{x}$ and $\widehat{y}$ trees are stored in temporary workspace, allowing the dense and low rank phases to overlap, requiring only a single stream synchronization between the dense phase and the final phase of the downsweep.

\subsection{Performance results}
\label{subsec_hmv_perf}
To demonstrate the performance of the HMV operation, we generated two families of hierarchical covariance matrices for a spatial Gaussian process with $n = 2^{14}\cdots2^{20}$ observation points placed on randomly perturbed 2D and 3D regular grids in $[0, 1]^2$ and $[0, 1]^3$ respectively. The covariance matrices were generated using an isotropic exponential kernel with correlation length 0.1 in 2D and 0.2 in 3D~\cite{ambi16} and are representative of hierarchical matrices that arise in several applications \cite{hackbusch15}.

The hierarchical matrices were generated \textit{ab initio} in the following way. The $n$ points are first partitioned into clusters using a KD-tree with a mean split, generating the index sets of the basis tree nodes. The basis and transfer matrices are then generated using Chebyshev interpolation \cite{borm05interpolation}. A dual traversal \cite{hackbush_2000,hackbush_2003} of the basis tree generates the quadtree described in Section \ref{sec:hmatrix_structure}, where the coupling matrices at the leaves are generated by evaluating the kernel at the interpolation points. The leaf size $m$ was set to 64, tuned to the P100 GPU, and a uniform rank of 64 was used for all levels of the matrix, corresponding to the use of $8\times8$ grids and $4\times4\times4$ Chebyshev grids in 2D and 3D respectively. The choice of the leaf size only influences performance and has no effect on the accuracy of the representation, since overall accuracy is limited by the low rank blocks. The resulting approximation error was less than $10^{-7}$ in 2D and less than $10^{-3}$ in 3D for all problem sizes, as measured by computing $\norm{Ax-A^\mathcal{H}x} / \norm{Ax}$ where $A$ is the exact (dense) covariance, $A^\mathcal{H}$ is its hierarchical representation, and $x$ is a randomly generated vector whose entries are sampled from a uniform $[0, 1]$ distribution. For the large problems, it is not possible to store the dense $A$ nor is it practical to perform the $\mathcal{O}(n^2)$ $A x$ product, and as a result we sampled 10\% of the rows and used the analytical expression of the matrix entries. While a rank of 64 may seem high for a leaf size of 64, it is often the case that the ranks at the leaves increase temporarily due to low rank updates that may be applied to the blocks of the matrix during hierarchical matrix operations. This rank will be reduced during compression and the effect on matrix vector multiplication performance is shown in Section \ref{sec:compperf}.

\begin{figure}
\begin{center}
\begin{tikzpicture}[spy using outlines={rectangle,orange,magnification=5,size=4.2cm, connect spies, 
						spy connection path={\draw[thick] (tikzspyonnode) -- (tikzspyinnode);},
	                    every spy on node/.append style={ultra thick}, 
						every spy in node/.append style={thick}}]
	\node [anchor=south west,inner sep=0] (image) at (0,0)
	             {\includegraphics[width=0.3\textwidth]{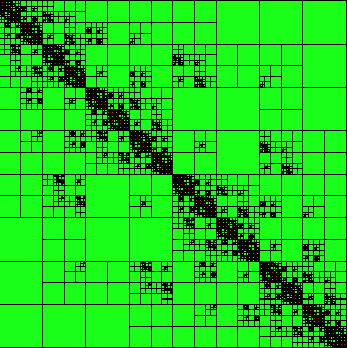} };		 
    \begin{scope}[x={(image.south east)},y={(image.north west)}]	
		\coordinate (p1) at (0.4,0.73);
		\coordinate (p2) at (2,0.5);
		\spy on (p1) in node [] at (p2); 
    \end{scope}	
\end{tikzpicture}
\end{center}
\begin{center}
\begin{tikzpicture}[spy using outlines={rectangle,orange,magnification=5,size=4.2cm, connect spies, 
						spy connection path={\draw[thick] (tikzspyonnode) -- (tikzspyinnode);},
	                    every spy on node/.append style={ultra thick}, 
						every spy in node/.append style={thick}}]
	\node [anchor=south west,inner sep=0] (image) at (0,0)
	             {\includegraphics[width=0.3\textwidth]{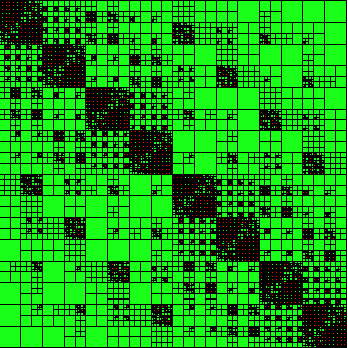} };		 
    \begin{scope}[x={(image.south east)},y={(image.north west)}]	
		\coordinate (p1) at (0.4,0.73);
		\coordinate (p2) at (2,0.5);
		\spy on (p1) in node [] at (p2); 
    \end{scope}	
\end{tikzpicture}
\end{center}
\caption{Structure of the 2D (top) and 3D (bottom) sample covariance matrices used in the examples for $n = 2^{14}$, with a zoom on portions of them. Notice that the 3D problem does not have as many large blocks that admit a low rank approximation as the 2D problem and therefore results in a representation that has higher memory demand for the same accuracy.} 
\label{fig:samples}
\end{figure}
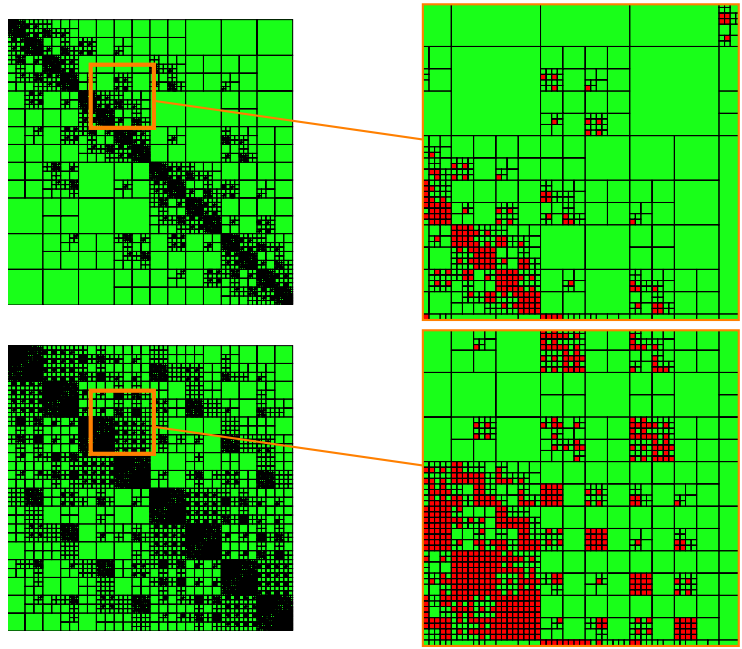

\begin{figure}[!ht]
	\centering
	\includegraphics[width=\textwidth]{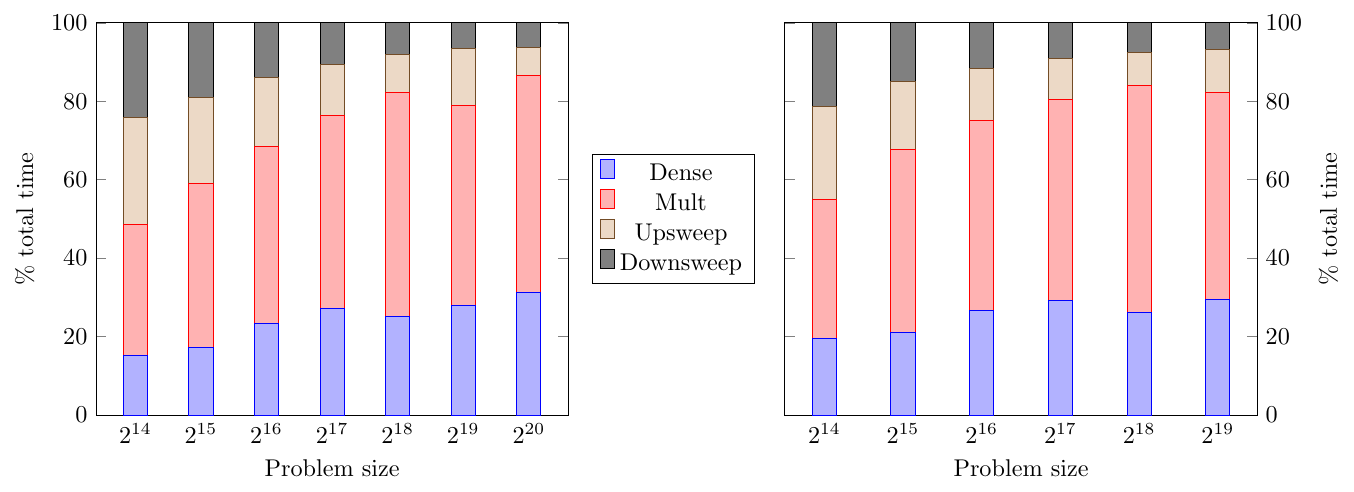}
	\caption{Breakdown of the HMV phases in percentages of total operation time in single (left) and double (right) precision on a P100 GPU showing that, for smaller problem sizes where the basis tree has very few levels, the upsweep and downsweep do not fully utilize the hardware.}
	\label{fig:hmv_profile}
\end{figure}

\begin{figure}[t]
	\begin{subfigure}[t]{0.45\textwidth}
	\centering
	\includegraphics[width=\textwidth]{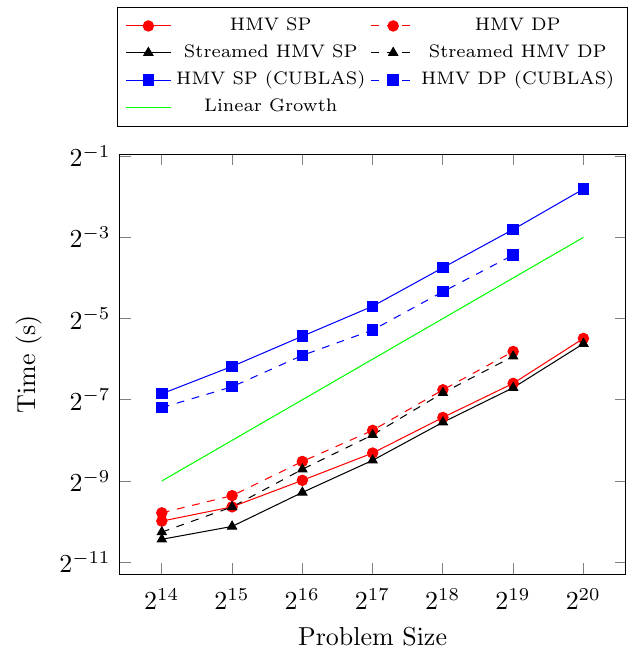}
	\caption{Time for the HMV in single and double precision for a 2D problem.}
	\label{fig:hmv_2d}
	\end{subfigure}
	\hfill	
	\begin{subfigure}[t]{0.45\textwidth}
	\centering
	\includegraphics[width=\textwidth]{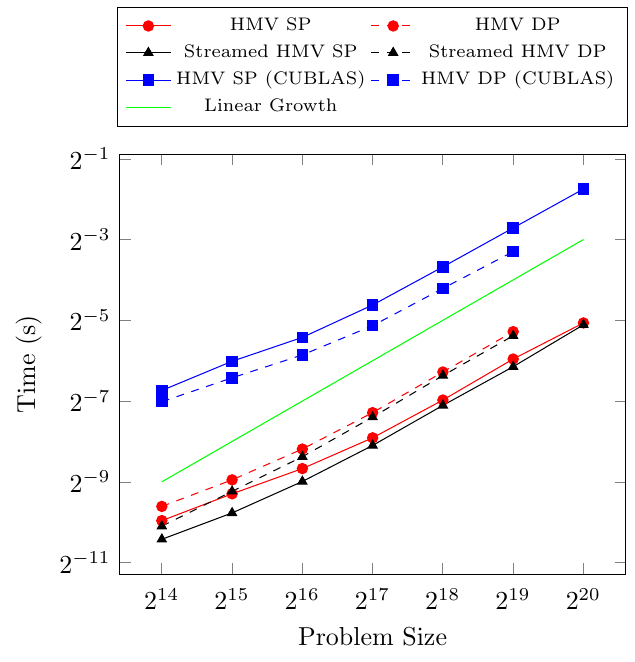}
	\caption{Time for the HMV in single and double precision for a 3D problem.}
	\label{fig:hmv_3d}
	\end{subfigure}
	\caption{Runtime of HMV on a single P100 GPU, showing asymptotically linear growth with problem size. Notice that the streamed version that allows overlapping between the dense and the low rank phases of HMV provides performance boost on small problems. On the larger problems, where the available bandwidth is saturated with the low rank data, the improvement due to overlapping is diminished.}
	\label{fig:hmv_timings}
\end{figure}	

\begin{figure}[t]
	\begin{subfigure}[t]{0.45\textwidth}
	\centering
	\includegraphics[width=\textwidth]{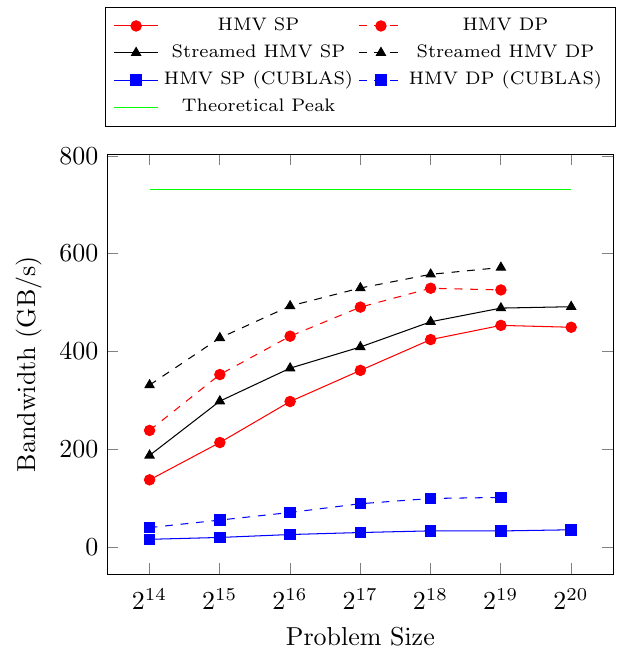}
	\caption{Achieved bandwidth for the HMV in single and double precision for a 2D problem.}
	\label{fig:hmv_bw_2d}
	\end{subfigure}
	\hfill	
	\begin{subfigure}[t]{0.45\textwidth}
	\centering
	\includegraphics[width=\textwidth]{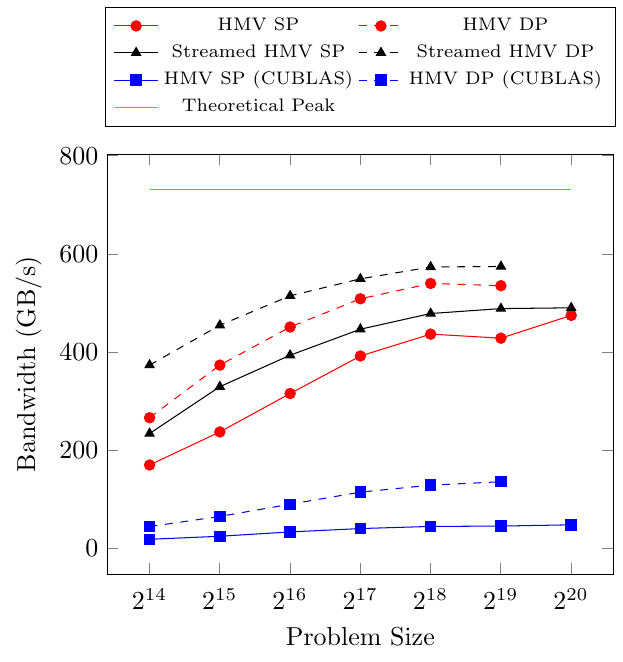}
	\caption{Achieved bandwidth for the HMV in single and double precision for a 3D problem.}
	\label{fig:hmv_bw_3d}
	\end{subfigure}
	\caption{Achieved bandwidth of HMV on a single P100 GPU with the streamed kernel achieving up to 78\% of the theoretical bandwidth peak of the GPU.}
	\label{fig:hmv_bandwidth}		
\end{figure}

For illustration, Figure~\ref{fig:samples} shows the structure of the 2D and 3D covariance matrices for the $n = 2^{14}$  problem size. The small dense blocks of size $64\times64$ are shown in red in the zoomed details. As expected, the 3D matrix has more dense blocks and its low rank trees are bushier than those of the 2D matrix which puts more pressure on memory (or alternatively permit lower accuracy for the same memory footprint as a 2D problem of the same size). The performance results shown here are for these matrices in their original analytically-derived hierarchical form. The algebraic compression discussed in section~\ref{sec:compression} will reduce the memory footprint of the low rank portions substantially and further reduce the HMV time, as will be shown in section~\ref{sec:compperf}. 

Figure \ref{fig:hmv_profile} shows the profile of the execution of the HMV of the 3D problems in single and double precision. The upsweep and downsweep phases show relatively poor performance for the smaller problems sizes, since the small batches generated for the upper levels of the trees do not provide enough work to saturate the GPU, leading to lower performance during those phases. The impact of the smaller higher levels is alleviated as the problem size increases, where the larger lower levels that can saturate the GPU dominate the workload. Kernel streaming as discussed in Section \ref{sec_hmv_stream} allows us to increase hardware usage during those phases.

Figures \ref{fig:hmv_2d} and \ref{fig:hmv_3d} show the execution time of the HMV in double and single precision for the 2D and 3D problems respectively. The streamed kernel shows up to $27\%$ performance improvement for the smaller problem sizes. We can also see the expected linear growth of the algorithm for all versions. We also compare the effect of using different batched kernels. All versions using the batched GEMV and block spMV kernels from~\cite{abdelfattah16a,abdelfattah16b} show significant improvements over the same algorithms implemented with the vendor provided routines in CUBLAS~\cite{cublas16} and CUSPARSE~\cite{cusparse16}. Since the matrix vector multiplication is a memory bound routine, we gauge the performance of the kernels by bandwidth. 

Since this operation is bandwidth limited, we compute performance as the total number of bytes transferred over total execution time. This includes all dense and coupling matrices in the matrix tree as well as the leaves and transfer matrices of the basis trees. Figures \ref{fig:hmv_bw_2d} and \ref{fig:hmv_bw_3d} show the achieved bandwidth of the various HMV kernels, with the streamed version achieving up to $78\%$ of the theoretical peak bandwidth of the P100 GPU. The improvement in achieved bandwidth over the non-streamed version that does not allow the overlap of the different portions of the computation is substantial for the small problem sizes. We note that performance was quite stable and reproducible between runs.


\section{Hierarchical matrix compression}
\label{sec:compression}

\subsection{Overview}
Compression is a core operation in hierarchical matrix algebra. For example, in the course of implementing BLAS3 operations, matrix blocks get added, generally producing increased apparent ranks. The matrix needs to be compressed in order to maintain the optimal complexity. The goal of compression is therefore to construct new nested row and columns bases in which the matrix blocks can be expressed more compactly, i.e., where blocks originally represented as $U_i S_{ij} V_j^T$  can be compressed into the form $U'_i S'_{ij} V_j^{\prime T}$ where the dimensions of $S'$ are smaller than the dimensions of $S$. The primary task here is to construct the common basis $U'_i$ in which all blocks of a given block row can be expressed, without incurring the quadratic cost that would be needed if a straight SVD of the whole row block is performed. The same goes for column blocks and $V'_j$ when the two bases $\mathcal{U}$ and $\mathcal{V}$ are different. Finally, once the compact and more efficient nested block row/column basis has been generated, the new $S'_{ij}$ for every matrix block is computed by a transformation of the form $T_{Ui} S_{ij} T_{Vj}^T$, leading to a smaller memory footprint and reduced hierarchical operation runtimes. The algorithms presented are adapted from~\cite{borm_2010} to fit the architecture of the GPU. 

\begin{figure}[t]
\centering
\includegraphics[width=0.4\textwidth]{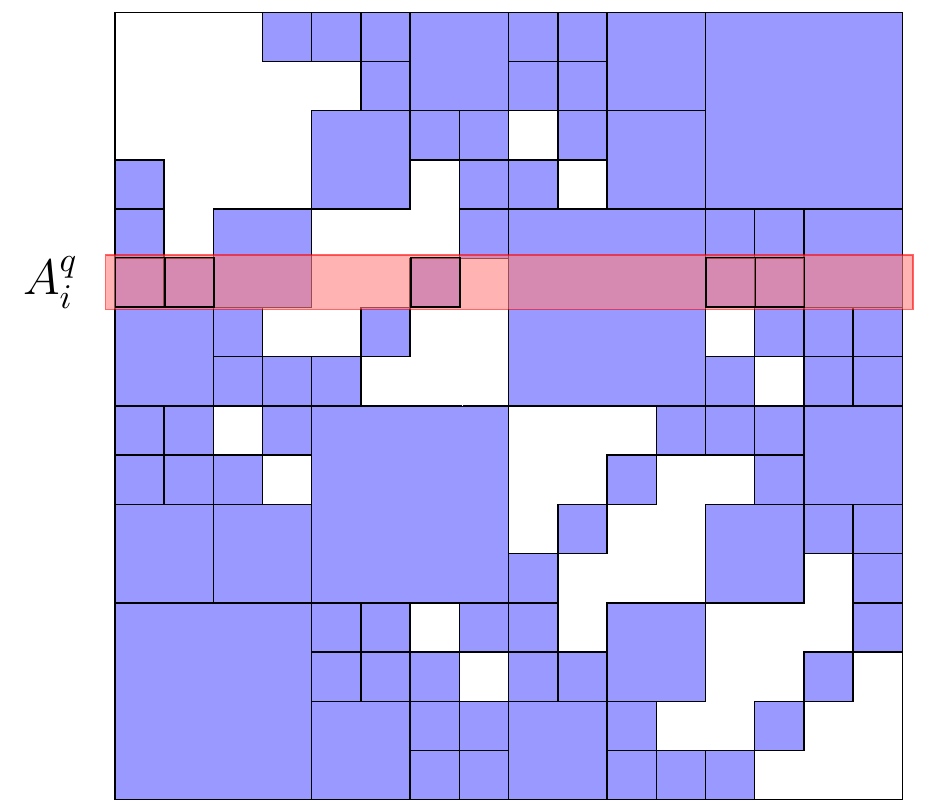}
\caption{Matrix block row $A_i^q$ of the low rank portion of the hierarchical matrix for the sixth row of the leaf level. $A_i^q$ includes 5 blocks at level $q$ and sub-blocks coming from higher level block rows.}
\label{fig:matrix_block_row}
\end{figure}

In order to introduce the somewhat involved algorithm, let's first consider how the new basis for a block row $A_i^q$ at the finest level $q$ would be generated. $A$ here denotes only the low rank portions of the hierarchical matrix, since the dense blocks are not compressed.  $A_i^q$ consists of $b$ low rank blocks expressed at level $q$ as $U_i S_{ij}V_{j}^T$ with $j=j_1 \cdots j_b$, and additional pieces representing the restriction of blocks from higher levels to their ``$t$'' rows as shown in Figure \ref{fig:matrix_block_row}. 
\begin{equation}
A_i^q = U_i^q \begin{bmatrix} \substack{\text{portions of} \\ \text{ancestors}} \quad  S_{ij_1}^q V_{j_1}^{q T} \cdots S_{ij_b}^q V_{j_b}^{q T} \end{bmatrix} 
      = U_i^q B_i^{qT} 	
\end{equation}
The optimal basis can be generated by computing the SVD of $U_i^q B_i^{qT}$, truncating it to the desired approximation, and using the truncated left singular vectors as the new basis $U^{\prime q}_i$. This would however require the expensive SVD of the $O(n)$-sized $B_i^q$. In order to avoid it, we would first compute the QR decomposition of $B_i^q$ and then perform the SVD with the small $R$ factor. 
\begin{equation}
A_i^q =  U_i^q {B_i^q}^T = U_i^q (Q_i^q R_i^q)^T  = \underbrace{U_i^q {R_i^q}^T}_{\text{new basis}}  {Q_i^q}^T	
\end{equation}
The optimal basis $U^{\prime q}_i$ is then simply the truncated left singular vectors of what might be thought of as a new weighted basis $U_i^q {R_i^q}^T$, and this finishes the process for level $q$. When we move to higher levels in the tree, we need to insure that the ${U'}$ bases remain nested. This requires additional singular value decompositions, but involving only small transfer matrices, as we go up $\mathcal{U}$ in an upsweep traversal described in Section \ref{sec:basis_truncation}. 

The task of computing $R_i^q$ of the QR decomposition of $B_i^q$ can be done efficiently by exploiting the nestedness of the bases.  Let us assume that the QR decomposition of $B_{i^+}^{q-1}$, the parent block $i^+$ at level $q-1$, is available as $Q_{i^+}^{q-1} R_{i^+}^{q-1}$. Then, 
\begin{equation}
\begin{split}
 A_i^q &=  \begin{bmatrix} \substack{\text{$i$-portion of} \\ U_{i^+}^{q-1} {B_{i^+}^{q-1}}^T}
          \quad U_i^q S_{ij_1}^q V_{j_1}^{q T} \cdots U_i^q S_{ij_b}^q V_{j_b}^{q T} \end{bmatrix} \\
	   & =  U_i^q \begin{bmatrix} E_i^q (Q_{i^+}^{q-1} R_{i^+}^{q-1})^T
          \quad S_{ij_1}^q V_{j_1}^{q T} \cdots S_{ij_b}^q V_{j_b}^{q T} \end{bmatrix} 
	   =  U_i^q B_i^{qT} 
\end{split}
\end{equation}
with $B_i^q$ conveniently expressible as:
\begin{equation}
	\label{eq:Btq}
B_i^q = \begin{bmatrix} Q_{i^+}^{q-1} \; R_{i^+}^{q-1} E_i^{q T} \\
	                    V_{j_1}^l S_{ij_1}^{q T} \\ \vdots \\ V_{j_b}^q S_{ij_b}^{q T}
        \end{bmatrix}
	  = \textbf{diag}(Q_{i^+}^{q-1}, V_{j_1}^q, \cdots, V_{j_b}^q)  
	    \begin{bmatrix} R_{i^+}^{q-1} E_i^{qT} \\ S_{ij_1}^{q T} \\ \vdots \\ S_{ij_b}^{q T}
	    \end{bmatrix}.
\end{equation}
Assuming the $V^q$ bases are orthogonal, the block diagonal matrix in Eq.~\ref{eq:Btq} is orthogonal and the QR  of $B_i^q$ simply reduces to the QR of the small stack at the end of Eq.~\ref{eq:Btq} which involves only $b+1$ blocks, each being a small $k \times k$ coupling/transfer matrix, and therefore can be done quite efficiently. Since this QR uses the $R^{q-1}$ matrix from level $q-1$, the overall computation starts from the root and goes down the tree computing all the $R^l_i$ matrices for all levels in a downsweep traversal. As with previous operations, all blocks at a given level can obviously be processed in parallel. We also observe here that the $Q$ factors are not needed and a specialized QR decomposition avoiding their expensive storage and computation improves the performance of the algorithm. 

Orthogonalizing the $\mathcal{V}$ basis tree can be done in a pre-processing phase. A basis is orthogonal if $V_j^{lT} V_j^l$ is the identity matrix for all levels $l$.  Orthogonalizing a basis involves performing QR on the finest level basis and then going up the tree to compute new transfer matrices that allow higher level nodes to satisfy the orthogonality condition. This is also done via additional QR operations involving the transfer matrices. 

In summary, the overall compression procedure consists of the following three computational steps:
\begin{itemize}
	\item An upsweep of the basis trees to orthogonalize them. This step uses a sequence of batched QR calls, one per level, to produce an orthogonal basis. This is described in Section \ref{sec:basis_orthogonalization}.
	\item A downsweep of the basis trees, using the coupling blocks $S_{ij}$, to construct the $R_i$ factors for the new bases. This step uses a sequence of batched QR kernel calls, one per level, on the stacks at the end of Eq.~\ref{eq:Btq}. This is described in Section \ref{sec:basis_generation}.
	\item An upsweep of the basis tree to truncate the new bases to the desired tolerance. This step uses a sequence of batched SVD calls, again one per level, on the $U_iR_i$ bases to produce the new optimal basis. The $S_{ij}$ blocks are then transformed into these bases via batched matrix multiplication operations. This is described in Section \ref{sec:basis_truncation}. 
\end{itemize}

\subsection{Basis orthogonalization}
\label{sec:basis_orthogonalization}

Orthogonalizing a nested basis tree $\mathcal{V}$ replaces it by a new nested basis where every node satisfies the orthogonality condition $V_j^{lT} V_j^l = I$. This is equivalent to the conditions that $V_j^{qT} V_j^q = I$ at the finest level $q$ and that the transfer matrices satisfy $\sum_{c} F_c^T F_c = I$ for all levels, where $c$ ranges over the two children of every node in the basis tree. Besides allowing the simplification in the $R$ computation algorithm, orthogonalizing the basis of an $\mathcal{H}$-matrix simplifies error computations and matrix projections. It is also structurally similar to the truncation algorithms described in Section \ref{sec:basis_truncation}. 

A by-product of orthogonalization is also a set of projection matrices that transform between the old basis and the new orthogonal one. These projection matrices are stored in a tree that shares the same structure as the basis tree, where processing each node of the basis tree produces a node in the projection tree $T_\mathcal{V}$. We assume here that $\mathcal{V} = \mathcal{U}$ and do not make a distinction between the row and column bases and drop the subscript $\mathcal{V}$ from $T$. If we denote by $Q^l_i$ the new orthogonal basis at level $l$, the original basis $V_j^l$ can be recovered as $Q_j^l T_j^l$. We will use this transformation to express the coupling matrices in the new orthogonal basis. 

As in the upsweep of the HMV algorithm, we perform the orthogonalization operation one level at a time, starting at the leaves and sweeping up the tree. Processing the leaves simply requires QR factorization of each leaf node where the $Q$ factor becomes the new orthogonal leaf and the $R$ factor is output into the leaf level of $T$. The inner nodes are expressed in terms of their children using the nested basis property and must be orthogonalized in a way that preserves this property. Given an inner node $j^+$ at level $l-1$ with children $j_1$ and $j_2$ at processed level $l$, we have
\begin{equation*}
V^{l-1}_{j^+} = \begin{bmatrix}
V^{l}_{j_1}  & 		\\
& V^{l}_{j_2} 
\end{bmatrix}\begin{bmatrix}
F_{j_1}\\
F_{j_2}
\end{bmatrix} = \begin{bmatrix}
Q^{l}_{j_1}  & 		\\
& Q^{l}_{j_2} 
\end{bmatrix}\begin{bmatrix}
T^{l}_{j_1} F_{j_1}\\
T^{l}_{j_2} F_{j_2}
\end{bmatrix} = \begin{bmatrix}
Q^{l}_{j_1}  & 		\\
& Q^{l}_{j_2} 
\end{bmatrix} Z
\end{equation*}

\begin{algorithm}[t]
\caption{GPU orthogonalization upsweep marshaling routine}
\label{alg:horthog_marshalupsweep}
\begin{algorithmic}[1]
\Procedure{marshalQRupsweep}{$Z, \widehat{T}^{(l)}, F^{(l)}$}
	\State $n_p$ = levelptr$[l-1]$
	\State $k_p$ = levelrank$[l-1]$
	\State $n_c$ = levelptr$[l]$
	\State $k_c$ = levelrank$[l]$	
	\ForAllp{$p = n_p \rightarrow n_c$} 
		\State $i = p - n_p$     					\Comment{{\small Batch index}}
		\State $c$ = head$[p]$ 
		\State $c_i = 0$
		\While {$c \neq \text{empty}$}
			\State $\batch{Z}[c - n_c] = $ ptr$\left( Z \right) + i \times 2 \times k_c \times k_p + c_i$  		 	\Comment{{\small Binary Tree}}
			\State $\batch{T}[c - n_c] = $ ptr$\left( \widehat{T}^{(l)} \right) + (c - n_c) \times k_c \times k_c$		\Comment{{\small Extract level pointer data}}
			\State $\batch{F}[c - n_c] = $ ptr$\left( F^{(l)} \right) + (c - n_c) \times k_c \times k_p$
			\State $c_i = c_i + k_c$
			\State $c = $ next$[c]$
		\EndWhile
	\EndForAllp
\EndProcedure
\end{algorithmic}
\end{algorithm}

\begin{figure}[b]
\centering
\includegraphics[width=\textwidth]{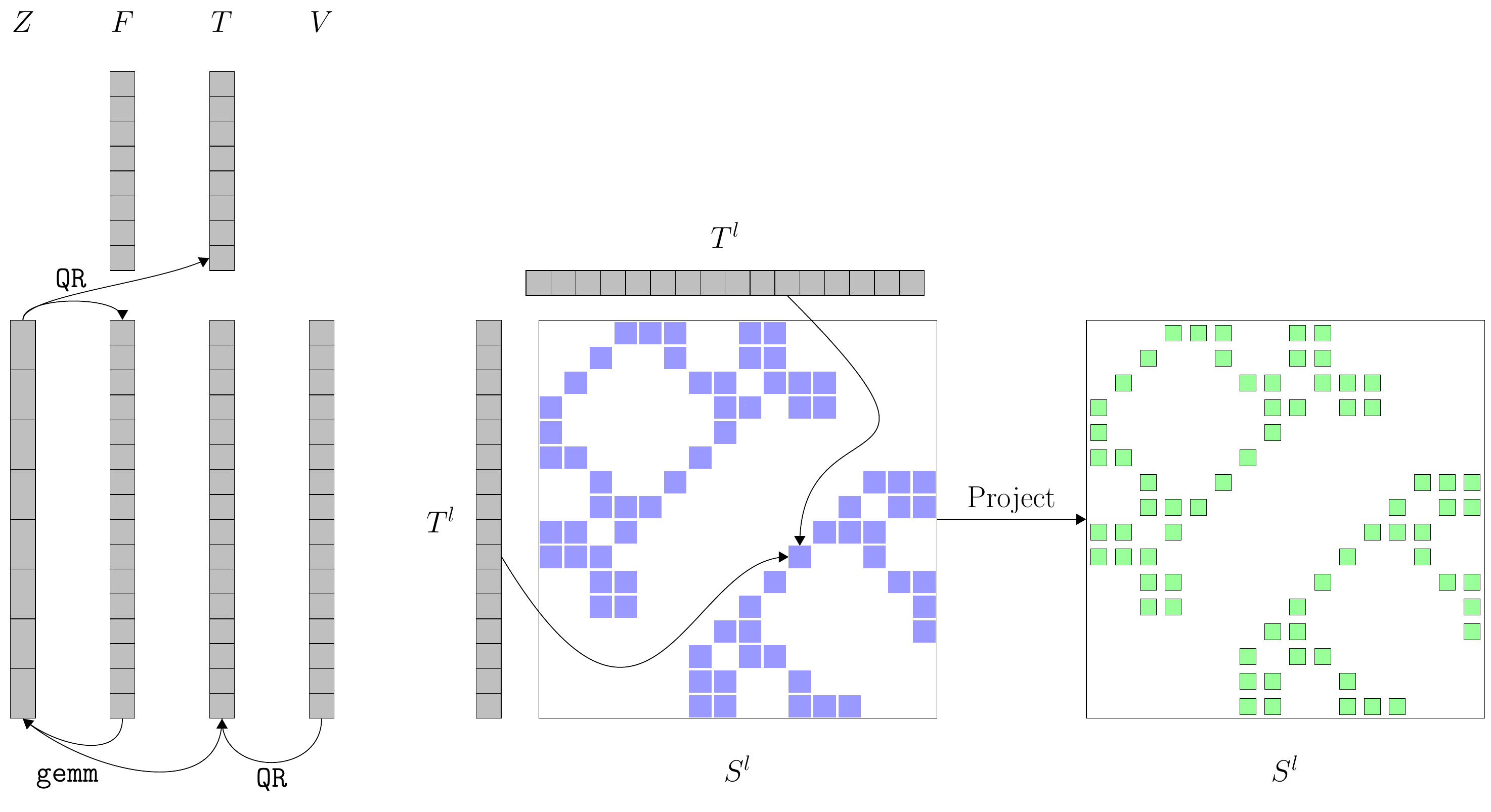}
\caption{Left: Basis orthogonalization starting from the leaves and sweeping up the tree to overwrite the basis with an orthogonal one and generate the projection tree $T$ (only a single level is depicted here). Right: Projection of the leaf level of coupling matrices into a new basis using a projection tree $T$. The new basis could be more compact, resulting in lower rank for the level.}
\label{fig:horthog_project}
\end{figure}

Forming the $2k^{l} \times k^{l-1}$ matrix $Z$, computing its QR factorization and using the two $k^{l} \times k^{l-1}$ blocks of the $Q$ factor as the new transfer matrices, gives us the new orthogonal inner nodes that satisfies the nested basis property. The $R$ factor is then output into level $l-1$ of the projection tree $T$. $Z$ is formed by first marshaling the operations based on the data in the projection tree and the basis tree transfer nodes and then operating on them using a batched matrix-matrix multiplication routine. $Z$ is then factorized using a batch QR factorization routine and the sub-blocks are copied back as the new transfer matrices using a marshaled batch copy operation. This leads us to Algorithm \ref{alg:horthog_marshalupsweep} for marshaling the orthogonalization upsweep operations and Algorithm \ref{alg:ortho} for computing the projection tree $T$ and the new orthogonal basis. The left side of Figure \ref{fig:horthog_project} depicts this operation for the binary basis tree.

\begin{algorithm}[t]
	\caption{Basis Orthogonalization}
	\label{alg:ortho}
	\begin{algorithmic}[1]
		\Procedure{QRupsweep}{$V$, $F$, $T$}
		\State $q$ = heightof$( T )$     \Comment{{\small \emph{leaf level, log(n/m)}}}
		\State $[ V, \, T^{q} ] $ = qrBatched$\left( \frac{n}{m}, V  \right)$ 
		\For{$l$ = $q \rightarrow {}1$} 	\Comment{{\small \emph{up the tree}}}
		\State $N$ = $n / m / 2^{q-l}$
		\State $[\batch{Z}, \batch{T}, \batch{F}]$ = marshalQRupsweep$( Z, T^{l}, F^{l})$
		\State gemmBatched$(N, \batch{Z}, \batch{T}, \batch{F})$ 
		\State $[\batch{Z}, \, T^{l-1}]$ = qrBatched$(N/2, \batch{Z})$
		\State $\batch{F}$ = marshalQRunstack$(\batch{Z}, F^{l})$
		\State $F^{l}$ = copyBatched$(\batch{F})$
		\EndFor
		\EndProcedure
	\end{algorithmic}
\end{algorithm}	

\begin{algorithm}[ht]
	\caption{Projection of Coupling Matrices}
	\label{alg:qrmult}
	\begin{algorithmic}[1]
	\Procedure{Projection}{$T$, $S$} 
		\ForAllp{ $l$ = $1 \rightarrow q$}
			\State $TS = \text{ws}(S^{l})$				\Comment{{\small \emph{Temporary workspace}}}
			\State $[\batch{ {T_U} }, \batch{S}, \batch{TS}, \batch{ {T_V} }]$ = marshalProjection$( T^{l}, S^{l}, TS )$
			\State nb = sizeof($S^{l}$)
			\State gemmBatched$(\text{nb}, \batch{TS}, \batch{ {T_U} }, \batch{S})$
			\State gemmBatched$(\text{nb}, S^{l}, \batch{TS},  \batch{ {T_V} }^T )$
		\EndForAllp
	\EndProcedure
	\end{algorithmic}
\end{algorithm}

Finally, the projection phase transforms the coupling matrices of each matrix block $A^{l}_{ij}$ at level $l$ using the projection matrices stored in $T^{l}$:
\begin{equation*}
A^{l}_{ij} = U^{l}_i S^{l}_{ij} {V^{l\,T}_j} = Q^{l}_i \left(T^{l}_i S^{l}_{ij} T^{l\,T}_j \right) {Q^{l\,T}_j}.
\end{equation*}
The new coupling matrices are obtained by left and right multiplications with the computed projection matrices independently at all levels. The operations are first marshaled by level and then executed using batch matrix-matrix multiplication routines. This operation is described in Algorithm \ref{alg:qrmult} and depicted in the right panel of Figure \ref{fig:horthog_project} for a single level of the matrix tree.

\subsection{Basis generation}
In this phase we construct a basis tree $R$ for the block rows of the matrix tree. This tree will have the same structure as the row basis tree. Every node $i$ at every level $l$ will store the matrix $R_i^l$ which will postmultiply the corresponding $U_i^l$ to produce the new basis that will be truncated. As described earlier, $R_i^l$ depends on matrix data coming from bocks at level $l$ whose row basis node is $U_i$ and from higher level blocks that also span the block row $i$. 

\begin{figure}[b]
\centering
\includegraphics[width=0.85\textwidth]{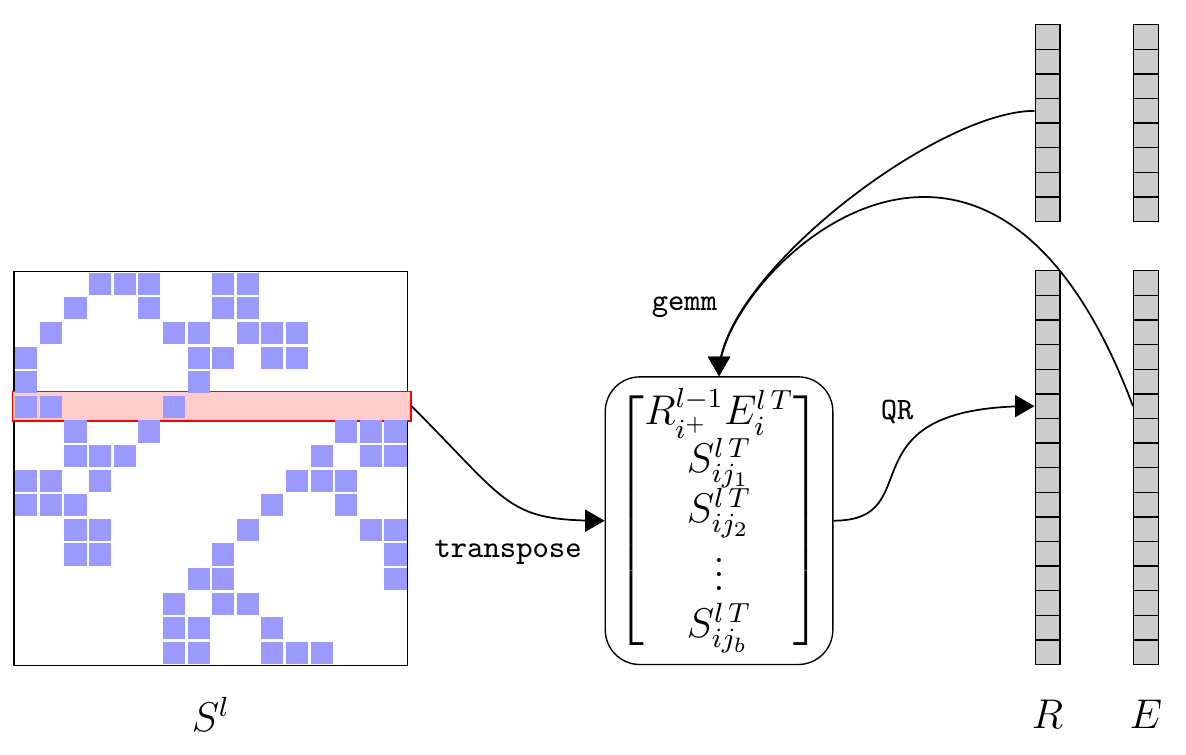}
\caption{Constructing $R_i^l$ from its parent level, the matrix tree row data, and transfer matrices.}
\label{fig:hbasisgen}
\end{figure}

Denoting the parent node of $i$ by $i^+$, the relevant block row $\bar{S}_i$ that enters the computation of $R_i^l$ is depicted in Figure \ref{fig:hbasisgen} and expressed as: 
\begin{equation*}
\bar{S}_i = \begin{bmatrix}
R_{i^+}^{l-1} E_i^{l\,T} \\
S_{ij_1}^{l\,T} \\
S_{ij_2}^{l\,T} \\
\vdots \\
S_{ij_b}^{l\,T} \\
\end{bmatrix}
\end{equation*}
where $b$ is the number of blocks in the block row at level $l$. The first block $R_{i^+}^{l-1} E_i^{lT}$ represents data coming from the levels of blocks above $l$. The node $R_i^{l}$ can then be computed as the $R$ factor of the QR factorization of $\bar{S}_i$. The tree is computed starting at the root of the matrix tree followed by a sweep down to the leaves. Marshaling the block row from the matrix tree data and the parent level of $R$ into a batch matrix-matrix multiplication and a batch transpose operation allows us to quickly form $\bar{S}_i$ for a level. This matrix is then fed into a batch QR factorization routine that does not form $Q$. The marshaling routine for this operation makes use of the generated BSR data of the level of the matrix tree and is described in Algorithm \ref{alg:marshalbasisgen}. Putting everything together leads us to Algorithm \ref{alg:basis} to form the tree $R$.

\begin{algorithm}[t]
\caption{GPU Basis Generation marshaling routine}
\label{alg:marshalbasisgen}
\begin{algorithmic}[1]
\Procedure{marshalBasisGen}{$R^{l-1}, E^l, S^l$}
	\State $n_r$ = levelrows$[l]$
	\State $k_p$ = levelrank$[l-1]$
	\State $k_c$ = levelrank$[l]$	
	\ForAllp{$r = 1 \rightarrow n_r$} 
		\State $b_s = $ rowPtr $[r]$     					
		\State $b_e = $ rowPtr $[r+1]$
		\State $p_r = $ parent $[r]$
		\State $\batch{RE}[r] = $ ptr $\left( \bar{S} \right) + r \times ld_S \times k_c$ \Comment{{\small $ld_S$ = rows of $\bar{S}$}}
		\State $\batch{R}[r] = $ ptr $\left( R^{l-1} \right) + p_r \times k_p \times k_p$
		\State $\batch{E}[r] = $ ptr $\left( E^l \right) + r \times k_p \times k_p$
		\For {$i = b_s \rightarrow b_e$} 
			\State $\batch{\bar{S}}[i] = $ ptr$\left( \bar{S} \right) + i \times ld_S \times k_c + k_p + (i - b_s) \times k_c$  
			\State $\batch{S}[i] = $ ptr$\left( S^{(l)} \right) + i \times k_c \times k_c$									
		\EndFor
	\EndForAllp
\EndProcedure
\end{algorithmic}
\end{algorithm}

\label{sec:basis_generation}
\begin{algorithm}[ht]
\caption{Basis Generation}
\label{alg:basis}
	\begin{algorithmic}[1]
	\Procedure{BasisGeneration}{$E$, $S$, $R$}
		\State $q$ = heightof$( R )$     \Comment{{\small \emph{leaf level, log(n/m)}}}
		\State $R^{0} = 0$
		\For{$l$ = $1 \rightarrow q$} 	\Comment{{\small \emph{down the matrix tree}}}
			\State $N$ = $n / m / 2^{q-l}$  \Comment{{\small \emph{number of block rows}}}
			\State nb = sizeof($S^l$)      \Comment{{\small \emph{ number of blocks in level $l$ }}}
			\State $[\batch{RE}, \batch{R}, \batch{E}, \batch{S}, \batch{\bar{S}}] $ = marshalBasisGen$( R^{l-1}, E^l, S^l)$
			\State gemmBatched$( N, \batch{RE}, \batch{R}, \batch{E}^T )$
			\State transposeBatched$(\text{nb}, \batch{S}, \batch{\bar{S}})]$
			\State $R^l$ = qrBatched*$(N, \bar{S} )$    \Comment{{\small \emph{ Q not computed }}}
		\EndFor
	\EndProcedure
	\end{algorithmic}
\end{algorithm}	

\subsection{Basis truncation}
\label{sec:basis_truncation}
\begin{algorithm}[hb]
	\caption{Basis Truncation}
	\label{alg:trunc}
	\begin{algorithmic}[1]
		\Procedure{SVDupsweep}{$U$, $E$, $R$, $T$, $\epsilon$}
		\State $q$ = heightof$( T )$     \Comment{{\small \emph{leaf level, log(n/m)}}}
		\State $W = $ gemmBatched$\left( \frac{n}{m}, U, R^{q} \right)$  \Comment{{\small \emph{new basis}}}
		\State $[ W, \widetilde{k}^{(q)} ] $ = svdBatched $\left( \frac{n}{m}, W , \epsilon \right)$ 
		    \Comment{{\small \emph{truncate the basis}}}
		\State $T^{q}$ = gemmBatched$\left( \frac{n}{m}, W^T, U \right)$ 
		    \Comment{{\small \emph{transformation between bases}}}
		\State $U = W$     \Comment{{\small \emph{replace basis by the new one}}}
		\For{$l$ = $q \rightarrow {}1$} 	\Comment{{\small \emph{up the tree}}}
		\State $N$ = $n / m / 2^{q-l}$
		\State $[\batch{T}, \batch{E}]$ = marshalSVDupsweep$(T^l, E^l)$
		\State $Z$ = gemmBatched$(N, \batch{T}, \batch{E})$ 
		\State $W$ = gemmBatched$(N/2, Z, R^{l-1})$
		\State $[W, \widetilde{k}^{l-1}]$ = svdBatched$(N/2, W, \epsilon)$
		\State $T^{l-1}$ = gemmBatched$\left( N/2, W^T, Z \right)$ 
		\State $\batch{F}$ = marshalSVDunstack$(W, F^l)$
		\State $F^l$ = copyBatched$(\batch{F})$
		\EndFor
		\EndProcedure
	\end{algorithmic}
\end{algorithm}	

Once the $R$ matrix of each block row is computed, we can generate the new compressed basis tree, which allows the ranks of the blocks at every level $l$ to decrease from $k^l$ to $\widetilde{k}^{l}$ while maintaining a prescribed accuracy $\epsilon$ in the approximation. This is the heart of algebraic compression. 

The truncation process is structurally similar to the upsweep of the orthogonalization in that processing the nodes produces projection matrices which are then used to sweep up the tree; however processing each node involves different computational kernels. For the leaf nodes $U_i^{q}$, we first use $R_i^q$ to produce a new basis node $W_i^{q} = U_i^{q} R_i^{qT}$. We then compute the singular value decomposition of  $W_i^{q}$ and produce the truncated basis by discarding singular vectors corresponding to values that are less than a threshold relative to the largest singular value. 

As we impose a constant rank per level, we use a fast reduction to compute the maximum truncated rank $\widetilde{k}^{q}$ for the given tolerance at the leaf level.  The truncated left singular vectors $Q_i^{q}$ will be the new compact basis node. Finally, the projection matrix into the new basis is computed as $T_i^{q} = Q_i^{qT} U_i^{q}$. 

Processing the inner nodes of the tree follows the same procedure as the orthogonalization: we form the $2\widetilde{k}^{l} \times k^{l-1}$ matrix $Z$ using the original transfer matrices and the projection matrices. We then compute a weighted $Z$ matrix, $W^{l}_i = Z_i R_i^l$, and proceed to compute its truncated singular value decomposition. The two $\widetilde{k}^{l} \times \widetilde{k}^{l-1}$ blocks of the truncated left singular vectors $Q_i^{l}$ will be the new transfer matrices for the truncated inner node, and the projection matrix is computed as $T_i^{l} = Q_i^{lT} Z_i$. The marshaling procedures are very similar to those of the orthogonalization with the addition of applying the factor $R$ to the original basis. The operations are then carried out by batch singular value decompositions and matrix-matrix multiplications. This leads us to Algorithm \ref{alg:trunc} for producing the new compact basis and the corresponding projection matrix tree $T$ given a relative threshold $\epsilon$ for truncation.

Finally, the projection of the coupling matrices is carried out in the same way as the projection phase of the orthogonalization procedure, using marshaled batch matrix-matrix multiplications with the projection matrix tree produced by Algorithm \ref{alg:trunc}.

\subsection{Performance results}
\label{sec:compperf}

We study the performance of the GPU compression procedure using the same two families of 2D and 3D covariance matrices described in Section \ref{subsec_hmv_perf} for HMV. The matrices were originally generated as hierarchical matrices in a generic polynomial basis.  As a result, their representation is not particularly memory efficient and algebraic compression can be therefore expected to produce new compressed hierarchical representation to reduce their memory footprint in an accuracy controllable way. 

\begin{table}[ht]
\centering
\def\arraystretch{1.2}
\begin{tabular}{|c|c|c|c|}
	\hline Problem size &
	$\frac{\norm{A^{\mathcal{H}}_2 - A^{\mathcal{H}}_1}_F}{\norm{A^{\mathcal{H}}_1}_F}$ &
	$\frac{\norm{Ax-A^{\mathcal{H}}_1x}}{\norm{Ax}}$ & 
	$\frac{\norm{Ax-A^{\mathcal{H}}_2x}}{\norm{Ax}}$ \\
	\hline \hline
	$2^{14}$ & $6.40 \times 10^{-8}$ & $3.33 \times 10^{-7}$ & $3.12 \times 10^{-7}$ \\
	$2^{15}$ & $9.85 \times 10^{-8}$ & $3.60 \times 10^{-7}$ & $3.36 \times 10^{-7}$ \\
	$2^{16}$ & $1.14 \times 10^{-7}$ & $3.47 \times 10^{-7}$ & $3.50 \times 10^{-7}$ \\
	$2^{17}$ & $1.52 \times 10^{-7}$ & $3.50 \times 10^{-7}$ & $3.58 \times 10^{-7}$ \\
	$2^{18}$ & $1.74 \times 10^{-7}$ & $3.49 \times 10^{-7}$ & $3.47 \times 10^{-7}$ \\
	$2^{19}$ & $2.19 \times 10^{-7}$ & $3.52 \times 10^{-7}$ & $3.48 \times 10^{-7}$ \\
	\hline
\end{tabular}
\caption{Compression errors for the 2D problem using a truncation threshold of $10^{-7}$.}
\label{fig:htrunc_err_2d}
\end{table}

\begin{table}[ht]
\centering
\def\arraystretch{1.2}
\begin{tabular}{|c|c|c|c|}
	\hline Problem size &
	$\frac{\norm{A^{\mathcal{H}}_2 - A^{\mathcal{H}}_1}_F}{\norm{A^{\mathcal{H}}_1}_F}$ &
	$\frac{\norm{Ax-A^{\mathcal{H}}_1x}}{\norm{Ax}}$ & 
	$\frac{\norm{Ax-A^{\mathcal{H}}_2x}}{\norm{Ax}}$ \\
	\hline \hline
	$2^{14}$ & $1.33 \times 10^{-3}$ & $9.10 \times 10^{-4}$ & $9.54 \times 10^{-4}$ \\
	$2^{15}$ & $1.75 \times 10^{-3}$ & $9.49 \times 10^{-4}$ & $1.03 \times 10^{-3}$ \\
	$2^{16}$ & $2.08 \times 10^{-3}$ & $9.19 \times 10^{-4}$ & $9.35 \times 10^{-4}$ \\
	$2^{17}$ & $2.35 \times 10^{-3}$ & $9.62 \times 10^{-4}$ & $9.70 \times 10^{-4}$ \\
	$2^{18}$ & $2.85 \times 10^{-3}$ & $9.78 \times 10^{-4}$ & $9.62 \times 10^{-4}$ \\
	$2^{19}$ & $2.83 \times 10^{-3}$ & $9.76 \times 10^{-4}$ & $9.68 \times 10^{-4}$ \\
	\hline
\end{tabular}
\caption{Compression errors for the 3D problem using a truncation threshold of $10^{-3}$.}
\label{fig:htrunc_err_3d}
\end{table}

Tables \ref{fig:htrunc_err_2d} and \ref{fig:htrunc_err_3d} show the compression errors for the 2D and 3D problems respectively. In this table, $A$ refers to the exact dense covariance (that is never formed but whose entries have analytical expressions from the underlying kernel), $A^\mathcal{H}_1$ refers to the hierarchal matrix approximation of the covariance generated using the generic Chebyshev polynomial basis described earlier, and 
$A^\mathcal{H}_2$ refers to the algebraically compressed covariance. The $A^\mathcal{H}_2$ matrices were generated using a truncation threshold of $10^{-7}$ for the 2D problems and $10^{-3}$ for the 3D problems. These thresholds were chosen to correspond to the errors that already existed in the $A^\mathcal{H}_1$ matrix approximation so that the algebraic compression does not introduce further approximation errors. Therefore the reduction in memory footprint comes purely from the generation of more efficient bases to represent the matrix. 

\begin{figure}
	\centering
	\includegraphics[width=\textwidth]{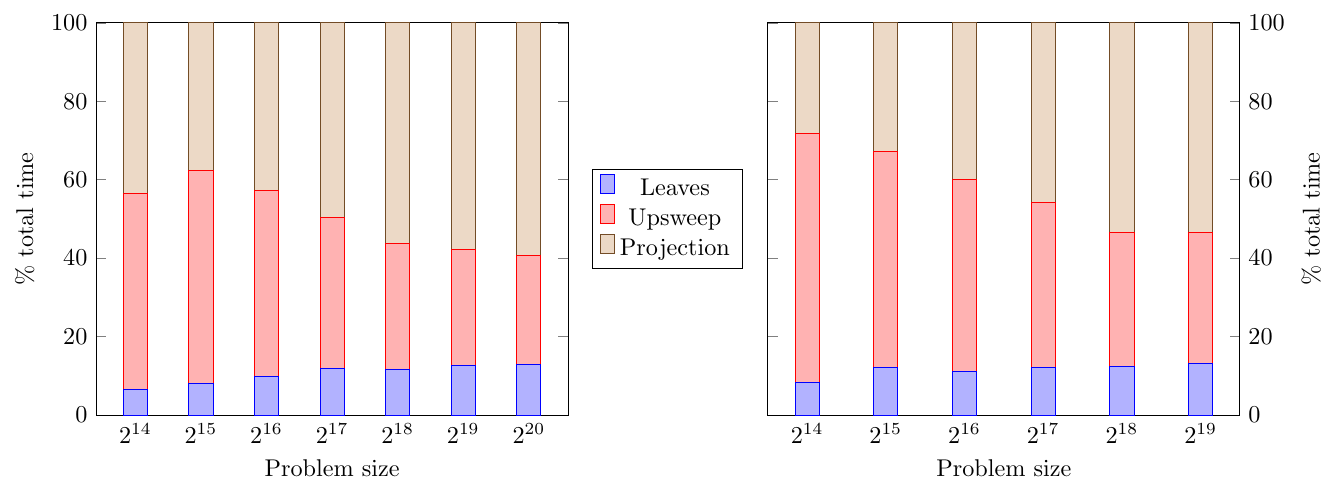}
	\caption{Breakdown of the orthogonalization phases in percentages of total operation time in single (left) and double (right) precision on a P100 GPU, showing that the projection phase dominates the runtime.}
	\label{fig:horthog_profile}

	\begin{subfigure}[t]{0.45\textwidth}
	\centering
	\includegraphics[width=\textwidth]{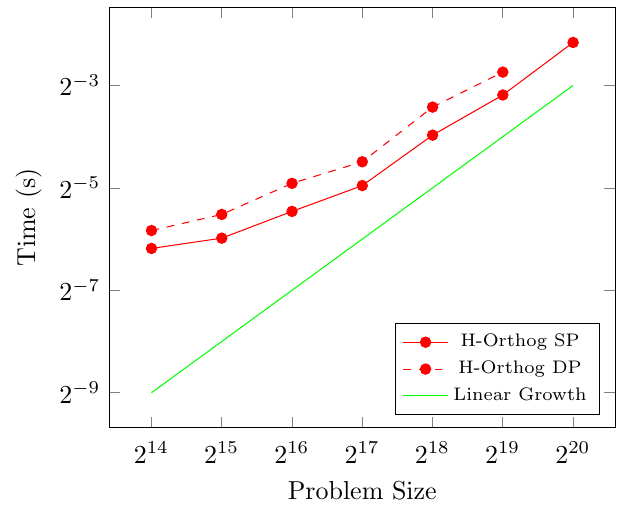}
	\caption{Time for the orthogonalization in single and double precision for a 2D problem.}
	\label{fig:horthog_2d}
	\end{subfigure}
	\hfill	
	\begin{subfigure}[t]{0.45\textwidth}
	\centering
	\includegraphics[width=\textwidth]{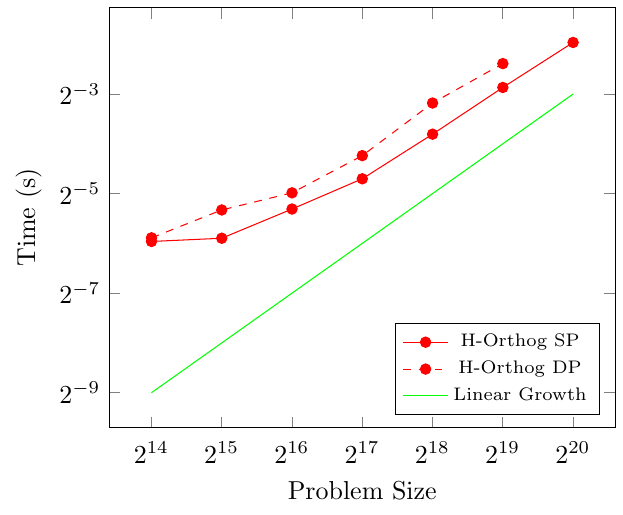}
	\caption{Time for the orthogonalization in single and double precision for a 3D problem.}
	\label{fig:horthog_3d}
	\end{subfigure}
	\begin{subfigure}[t]{0.45\textwidth}
	\centering
	\includegraphics[width=\textwidth]{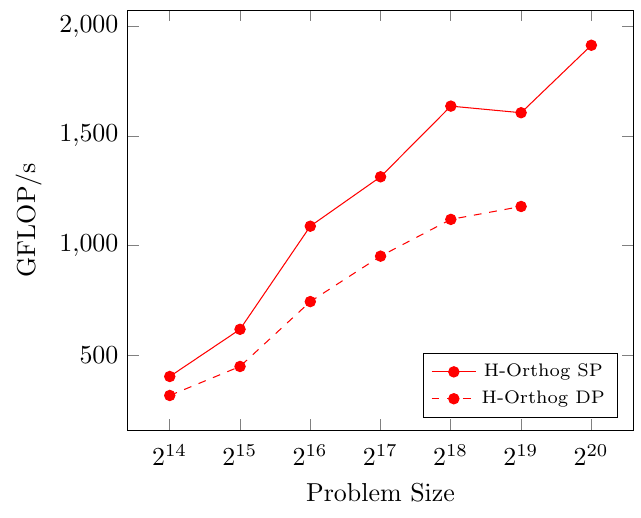}
	\caption{Performance of orthogonalization in single and double precision for a 2D problem.}
	\label{fig:horthog_perf_2d}
	\end{subfigure}
	\hfill	
	\begin{subfigure}[t]{0.45\textwidth}
	\centering
	\includegraphics[width=\textwidth]{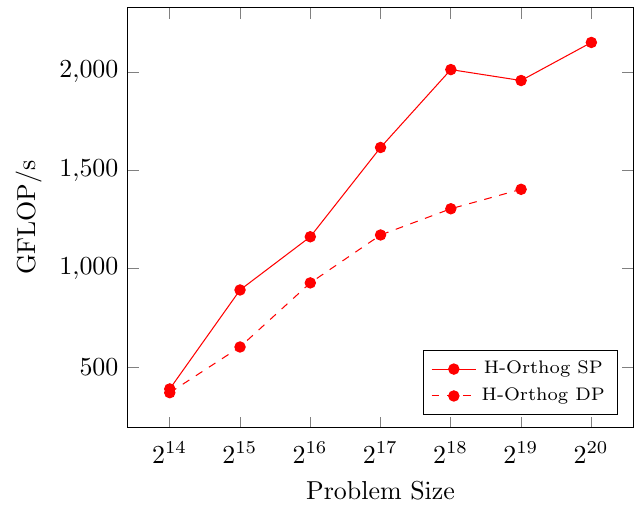}
	\caption{Performance of orthogonalization in single and double precision for a 3D problem.}
	\label{fig:horthog_perf_3d}
	\end{subfigure}
	\caption{Runtime and achieved performance of the orthogonalization on a single P100 GPU. Note the asymptotic linear growth with problem size.  Double and single precision times are closer to each other than expected due to the performance of the CUBLAS batched \texttt{gemm} routines.} 
	\label{fig:horthog_perf}		
\end{figure}

\begin{figure}
	\centering
	\includegraphics[width=\textwidth]{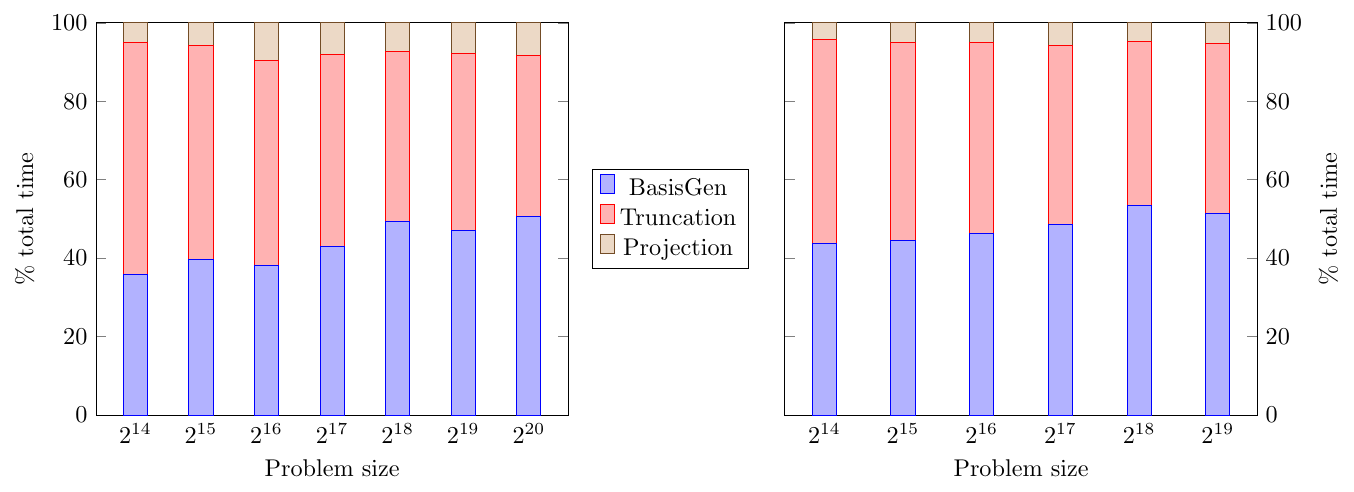}
	\caption{Breakdown of the compression phases in percentages of total operation time in single (left) and double (right) precision on a P100 GPU, showing that the truncation and basis generation phases dominate the computation.}
	\label{fig:htrunc_profile}

	\begin{subfigure}[t]{0.45\textwidth}
	\centering
	\includegraphics[width=\textwidth]{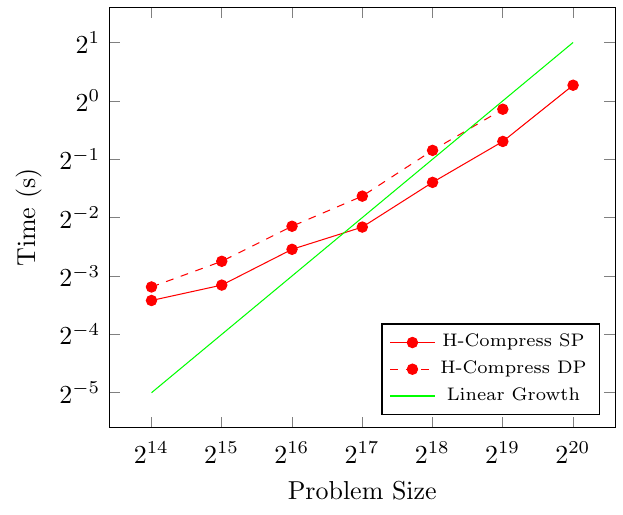}
	\caption{Time for the compression in single and double precision for a 2D problem.}
	\label{fig:htrunc_2d}
	\end{subfigure}
	\hfill	
	\begin{subfigure}[t]{0.45\textwidth}
	\centering
	\includegraphics[width=\textwidth]{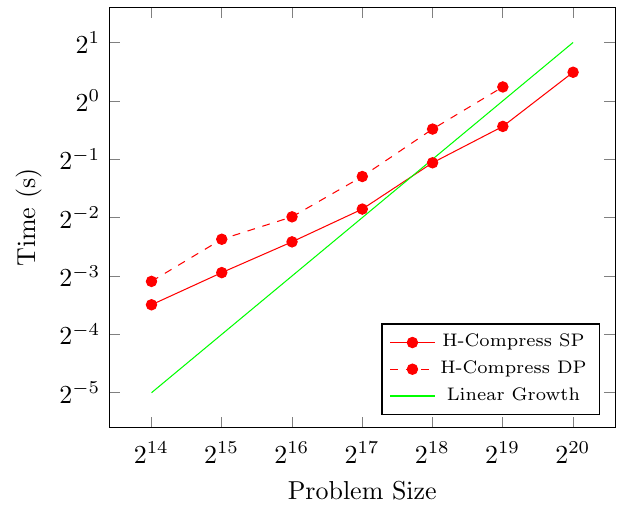}
	\caption{Time for the compression in single and double precision for a 3D problem.}
	\label{fig:htrunc_3d}
	\end{subfigure}
	
	\begin{subfigure}[t]{0.45\textwidth}
	\centering
	\includegraphics[width=\textwidth]{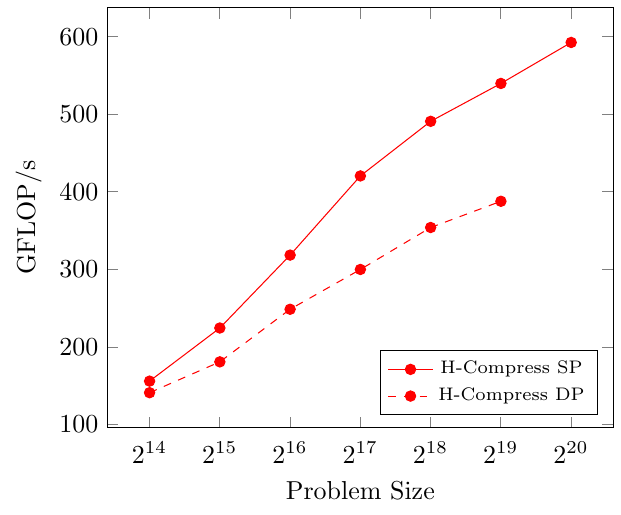}
	\caption{Performance of compression in single and double precision for a 2D problem.}
	\label{fig:htrunc_perf_2d}
	\end{subfigure}
	\hfill	
	\begin{subfigure}[t]{0.45\textwidth}
	\centering
	\includegraphics[width=\textwidth]{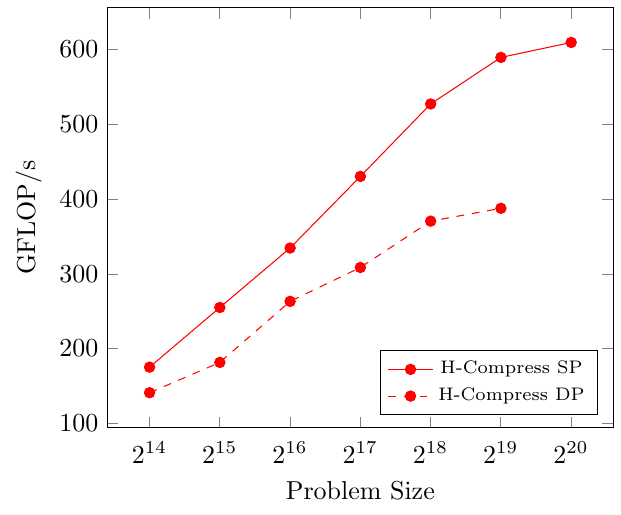}
	\caption{Performance of compression in single and double precision for a 3D problem.}
	\label{fig:htrunc_perf_3d}
	\end{subfigure}
	\caption{Runtime and achieved performance of compression on a single P100 GPU, showing asymptotically linear growth with problem size.}
	\label{fig:htrunc_perf}		
\end{figure}

The first column of Tables \ref{fig:htrunc_err_2d} and \ref{fig:htrunc_err_3d} shows the relative error between the original hierarchical matrix $A^\mathcal{H}_1$ and the compressed matrix $A^\mathcal{H}_2$ in the Frobenius norm. As expected, these errors are on the order of the truncation threshold used in the compression. We note here that the Frobenius norm error is computed, quite inexpensively, in the course of the truncation operation and does not require a separate post-processing operation. This is a useful feature for applications that require adaptive tolerances and fine error control inside chains of hierarchical operations. The second and third columns show the relative error pre- and post-compression respectively, measured in the 2-norm $\norm{Ax - A^\mathcal{H}x} / \norm{Ax}$ where $x$ is a random vector whose entries are uniformly distributed. For the large problems where it is too expensive to compute $Ax$, we randomly sampled $10\%$ of the rows and scaled the resulting error. As expected, comparing the second and third columns shows that compression with the appropriate thresholds had little to no effect on the accuracy of the resulting matrix. 

\begin{figure}
	\begin{subfigure}[t]{0.45\textwidth}
	\centering
	\includegraphics[width=\textwidth]{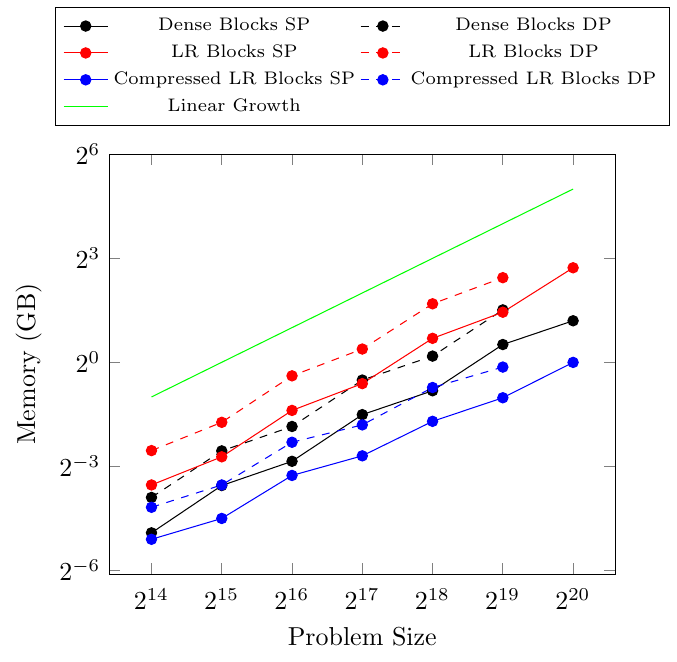}
	\caption{Effect of the compression on memory for the 2D problem with a truncation threshold of $10^{-7}$.}
	\label{fig:hmem_2d}
	\end{subfigure}
	\hfill	
	\begin{subfigure}[t]{0.45\textwidth}
	\centering
	\includegraphics[width=\textwidth]{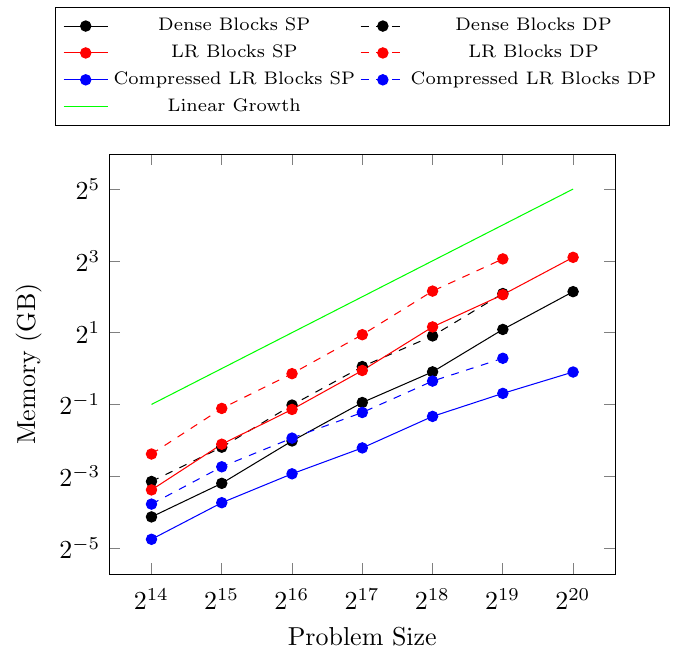}
	\caption{Effect of the compression on memory for the 3D problem with a truncation threshold of $10^{-3}$.}
	\label{fig:hmem_3d}
	\end{subfigure}
	\label{fig:hmem}
	
	\begin{subfigure}[t]{0.45\textwidth}
	\centering
	\includegraphics[width=\textwidth]{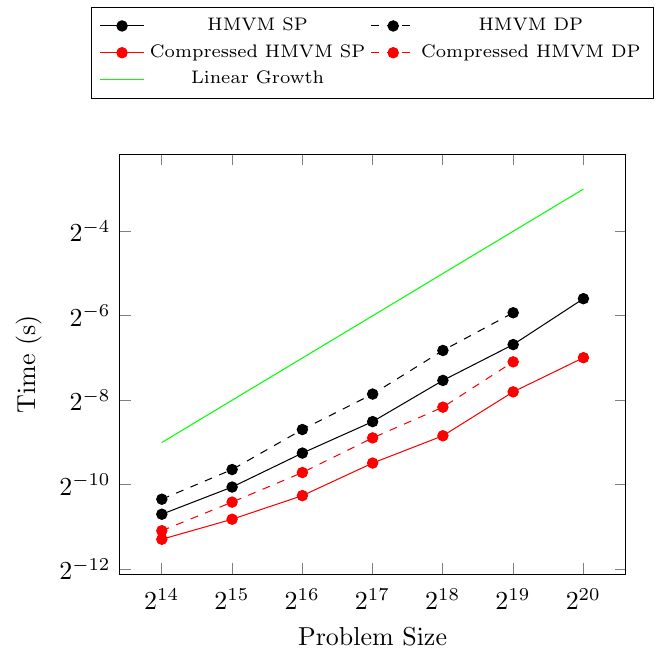}
	\caption{Effect of the compression on HMV for the 2D problem.}
	\label{fig:htrunc_hmv2d}
	\end{subfigure}
	\hfill	
	\begin{subfigure}[t]{0.45\textwidth}
	\centering
	\includegraphics[width=\textwidth]{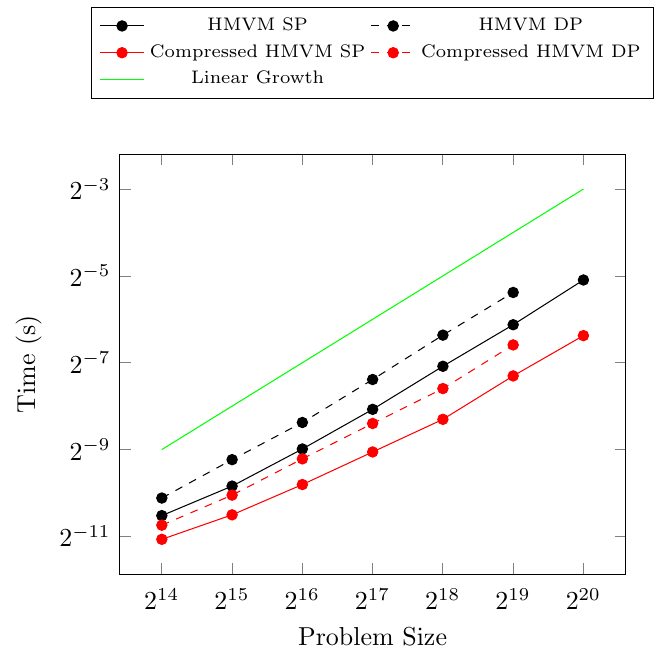}
	\caption{Effect of the compression on HMV for the 3D problem.}
	\label{fig:htrunc_hmv3d}
	\end{subfigure}
	\caption{Effect of the compression on memory and runtime for the 2D and 3D problems, showing significant memory savings for the low rank portion and the HMV time.}
	\label{fig:htrunc_hmv}
\end{figure}

We first profile and measure the performance of the orthogonalization kernel. Figure \ref{fig:horthog_profile} shows the percentage of total operation time spent in each of the three phases of the orthogonalization in single and double precision. It is easy to see that the projection phase dominates the runtime for the larger problem sizes, making the computational kernel at the core of this operation, the batched matrix-matrix multiplication, the main performance limiter. On the P100 GPU and for small matrix sizes, the CUBLAS batched \texttt{gemm} routines we use show lackluster performance and stand to be improved; the single precision performance is actually quite close to that of the double precision. We see this effect in the runtimes of the orthogonalization in Figures \ref{fig:horthog_2d} and \ref{fig:horthog_3d}; however, the expected linear growth of the algorithm remains. As the kernels involved in this computation are compute bound, Figures \ref{fig:horthog_perf_2d} and \ref{fig:horthog_perf_3d} show the performance in single and double precision for the 2D and 3D problems in GFLOP/s. Overall however, the orthogonalization phase represents a relatively small percentage (10--15\%) of the total compression time. 

Figure \ref{fig:htrunc_profile} shows the runtime profile of the compression in single and double precision where both truncation and basis generation phases are dominant due to the relatively costly SVD operations on the basis nodes and the QR decompositions on the coupling matrix data, especially when compared to the matrix-matrix multiplications of the projection phase. Figures \ref{fig:htrunc_2d} and \ref{fig:htrunc_3d} show the asymptotic growth of the compression algorithm. Since this is also a compute bound algorithm, we show the performance of the compression in GFLOP/s as the total number of operations executed by each batched kernel over total execution time in Figures \ref{fig:htrunc_perf_2d} and \ref{fig:htrunc_perf_3d}. 

Figures \ref{fig:hmem_2d} and \ref{fig:hmem_3d} show the significant memory savings achieved by the algebraic compression procedure. The compression also has a positive effect on the runtime of the hierarchical matrix arithmetic as shown in the improved runtimes of the matrix-vector operation in Figures \ref{fig:htrunc_hmv2d} and \ref{fig:htrunc_hmv3d}.


\section{Summary and Conclusions}

Hierarchical matrices provide memory-efficient approximations for many of the dense matrices that appear in a variety of contexts in scientific computing. In their $\mathcal{H}^2$ form, which may be viewed as an algebraic generalization of fast multipole methods, they have asymptotically optimal memory requirements and can store approximations of dense matrices in $\mathcal{O}(n)$ units of storage. This optimal memory footprint makes them particularly useful representations on modern hardware architectures, which are characterized by their limited memory relative to the raw arithmetic performance of their cores, such as GPUs and manycore processors.  

The objective of this work is to develop algorithms for operating on $\mathcal{H}^2$ matrices, that expose the fine grained parallelism to allow for efficient GPU execution. We describe algorithms and high performance implementations of hierarchical matrix vector multiplication as well as hierarchical matrix compression, a key component to efficient $\mathcal{H}^2-$matrix arithmetic, and show that the computations are amenable to highly efficient processing on GPUs.  By flattening the representation trees of a hierarchical matrix, we can efficiently marshal the operations for parallel execution using high performance batched kernels, and cleanly separate the linear algebra from the tree operations. 

Operating directly on the compressed representation, we demonstrate that the matrix vector operation can be completed in under $29$ms at $78\%$ of the theoretical peak bandwidth on the P100 GPU for a problem size of over a million involving a representative covariance matrix arising in 3D spatial statistics. The compression of a matrix from an initial hierarchical representation generated from an analytical kernel to an optimal algebraically-compressed one that preserves the original accuracy, can be done in under $1.7$s for the million sized 3D problem and executes at nearly $850$ GFLOPS/s on the P100 GPU, including a basis orthogonalization phase that executes at more than $2,000$ GFLOPS/s and basis generation and truncation phases at over $600$ GFLOPS. 

With the core compression algorithm in place, our future work will tackle BLAS3 operations on hierarchical matrices, including matrix-matrix multiplication and matrix inversion on GPUS, and explore applications of hierarchical matrices that can benefit from the high performance of GPUs. 



\bibliographystyle{ieeetran}
\bibliography{toms}

\begin{thebibliography}{10}
\providecommand{\url}[1]{#1}
\csname url@samestyle\endcsname
\providecommand{\newblock}{\relax}
\providecommand{\bibinfo}[2]{#2}
\providecommand{\BIBentrySTDinterwordspacing}{\spaceskip=0pt\relax}
\providecommand{\BIBentryALTinterwordstretchfactor}{4}
\providecommand{\BIBentryALTinterwordspacing}{\spaceskip=\fontdimen2\font plus
\BIBentryALTinterwordstretchfactor\fontdimen3\font minus
  \fontdimen4\font\relax}
\providecommand{\BIBforeignlanguage}[2]{{%
\expandafter\ifx\csname l@#1\endcsname\relax
\typeout{** WARNING: IEEEtran.bst: No hyphenation pattern has been}%
\typeout{** loaded for the language `#1'. Using the pattern for}%
\typeout{** the default language instead.}%
\else
\language=\csname l@#1\endcsname
\fi
#2}}
\providecommand{\BIBdecl}{\relax}
\BIBdecl

\bibitem{hlibpro}
\BIBentryALTinterwordspacing
R.~Kriemann, \emph{$\mathcal{H}$-lib pro}, 2016. [Online]. Available:
  \url{www.hlibpro.com}
\BIBentrySTDinterwordspacing

\bibitem{kriemann13}
------, ``{H-LU factorization on many-core systems},'' \emph{Computing and
  Visualization in Science}, vol.~16, no.~3, pp. 105--117, 2013.

\bibitem{borm15}
S.~B{\"{o}}rm and S.~Christophersen, ``{Approximation of {BEM} matrices using
  GPGPUs},'' \emph{arXiv}, vol. {abs/1510.07244}, 2015.

\bibitem{pzaspel17}
\BIBentryALTinterwordspacing
P.~Zaspel, ``Algorithmic patterns for h-matrices on many-core processors,''
  \emph{CoRR}, vol. abs/1708.09707, 2017. [Online]. Available:
  \url{http://arxiv.org/abs/1708.09707}
\BIBentrySTDinterwordspacing

\bibitem{filippone17}
S.~Filippone, V.~Cardellini, D.~Barbieri, and A.~Fanfarillo, ``Sparse
  matrix-vector multiplication on gpgpus,'' \emph{ACM Transactions on
  Mathematical Software}, vol.~43, no.~4, pp. 30:1--30:49, 2017.

\bibitem{bell12}
N.~Bell, S.~Dalton, and L.~Olson, ``Exposing fine-grained parallelism in
  algebraic multigrid methods,'' \emph{SIAM Journal on Scientific Computing},
  vol.~34, no.~4, pp. 123--152, 2012.

\bibitem{bell09}
N.~Bell and M.~Garland, ``Implementing sparse matrix-vector multiplication on
  throughput-oriented processors,'' in \emph{SC '09: Proceedings of the
  Conference on High Performance Computing Networking, Storage and Analysis},
  2009, pp. 1--11.

\bibitem{merrill16}
D.~Merrill and M.~Garland, ``Merge-based parallel sparse matrix-vector
  multiplication,'' in \emph{SC16: International Conference for High
  Performance Computing, Networking, Storage and Analysis}, 2016, pp. 678--689.

\bibitem{stream}
T.~Deakin, J.~Price, M.~Martineau, and S.~McIntosh-Smith, \emph{{GPU-STREAM
  v2.0}: Benchmarking the Achievable Memory Bandwidth of Many-Core Processors
  Across Diverse Parallel Programming Models}, ser. ISC High Performance 2016.
  Lecture Notes in Computer Science.\hskip 1em plus 0.5em minus 0.4em\relax
  Springer, 2016, vol. 9945, pp. 489--507.

\bibitem{hackbusch15}
W.~Hackbusch, \emph{Hierarchical Matrices: Algorithms and Analysis}.\hskip 1em
  plus 0.5em minus 0.4em\relax Springer, 2015.

\bibitem{ballani16}
J.~Ballani and D.~Kressner, \emph{Matrices with Hierarchical Low-Rank
  Structures}.\hskip 1em plus 0.5em minus 0.4em\relax Cham: Springer
  International Publishing, 2016, pp. 161--209.

\bibitem{hackbusch99}
W.~Hackbusch, ``A sparse matrix arithmetic based on {$\mathcal{H}$}-matrices.
  part i: Introduction to {$\mathcal{H}$}-matrices,'' \emph{Computing},
  vol.~62, no.~2, pp. 89--108, 1999.

\bibitem{borm02}
W.~Hackbusch and S.~B{\"o}rm, ``Data-sparse approximation by adaptive
  h2-matrices,'' \emph{Computing}, vol.~69, no.~1, pp. 1--35, 2002.

\bibitem{borm05}
S.~B{\"o}rm, ``Approximation of integral operators by matrices with adaptive
  bases,'' \emph{Computing}, vol.~74, no.~3, pp. 249--271, 2005.

\bibitem{borm2010}
------, ``Approximation of solution operators of elliptic partial differential
  equations by $\mathcal{H}$- and $\mathcal{H}^2$-matrices,'' \emph{Numerische
  Mathematik}, vol. 115, no.~2, pp. 165--193, 2010.

\bibitem{grasedyck2003}
L.~Grasedyck, W.~Hackbusch, and B.~N. Khoromskij, ``Solution of large scale
  algebraic matrix riccati equations by use of hierarchical matrices,''
  \emph{Computing}, vol.~70, no.~2, pp. 121--165, 2003.

\bibitem{grasedyck2009}
\BIBentryALTinterwordspacing
L.~Grasedyck, R.~Kriemann, and S.~Le~Borne, ``Domain decomposition based
  $\mathcal{H}$-lu preconditioning,'' \emph{Numerische Mathematik}, vol. 112,
  no.~4, pp. 565--600, 2009. [Online]. Available:
  \url{http://dx.doi.org/10.1007/s00211-009-0218-6}
\BIBentrySTDinterwordspacing

\bibitem{hackbusch04}
W.~Hackbusch, B.~N. Khoromskij, and R.~Kriemann, ``Hierarchical matrices based
  on a weak admissibility criterion,'' \emph{Computing}, vol.~73, no.~3, pp.
  207--243, Oct 2004.

\bibitem{xia10}
J.~Xia, S.~Chandrasekaran, M.~Gu, and X.~S. Li, ``Fast algorithms for
  hierarchically semiseparable matrices,'' \emph{Numerical Linear Algebra with
  Applications}, vol.~17, no.~6, pp. 953--976, 2010.

\bibitem{gillman2012direct}
A.~Gillman, P.~M. Young, and P.-G. Martinsson, ``A direct solver with o (n)
  complexity for integral equations on one-dimensional domains,''
  \emph{Frontiers of Mathematics in China}, vol.~7, no.~2, pp. 217--247, 2012.

\bibitem{Martinsson2013}
P.-G. Martinsson, ``A direct solver for variable coefficient elliptic pdes
  discretized via a composite spectral collocation method,'' \emph{Journal of
  Computational Physics}, vol. 242, pp. 460 -- 479, 2013.

\bibitem{ambikasaran2013n}
S.~Ambikasaran and E.~Darve, ``An o(n log n) fast direct solver for partial
  hierarchically semi-separable matrices,'' \emph{Journal of Scientific
  Computing}, vol.~57, no.~3, pp. 477--501, 2013.

\bibitem{batch_haidar}
A.~Haidar, T.~Dong, P.~Luszczek, S.~Tomov, and J.~Dongarra, ``Optimization for
  performance and energy for batched matrix computations on {GPU}s,'' in
  \emph{Proceedings of the 8th Workshop on General Purpose Processing Using
  GPUs}, ser. GPGPU-8.\hskip 1em plus 0.5em minus 0.4em\relax New York, NY,
  USA: ACM, 2015, pp. 59--69.

\bibitem{bell2011thrust}
N.~Bell and J.~Hoberock, ``Thrust: A productivity-oriented library for cuda,''
  in \emph{GPU computing gems Jade edition}.\hskip 1em plus 0.5em minus
  0.4em\relax Elsevier, 2011, pp. 359--371.

\bibitem{BOUKARAM2017}
\BIBentryALTinterwordspacing
W.~H. Boukaram, G.~Turkiyyah, H.~Ltaief, and D.~E. Keyes, ``Batched qr and svd
  algorithms on gpus with applications in hierarchical matrix compression,''
  \emph{Parallel Computing}, 2017. [Online]. Available:
  \url{http://www.sciencedirect.com/science/article/pii/S0167819117301461}
\BIBentrySTDinterwordspacing

\bibitem{cublas16}
NVIDIA, \emph{{cuBLAS} Library Documentation (v8.0)}, 2016.

\bibitem{dongarra14}
{Dongarra et al.}, \emph{{Accelerating Numerical Dense Linear Algebra
  Calculations with GPUs}}.\hskip 1em plus 0.5em minus 0.4em\relax Springer,
  2014, ch. {Numerical Computations with GPUs}, pp. 1--26.

\bibitem{sfi14}
R.~Yokota, G.~Turkiyyah, and D.~Keyes, ``{Communication Complexity of the Fast
  Multipole Method and Its Algebraic Variants},'' \emph{Supercomputing
  Frontiers and Innovations}, vol.~1, no.~1, pp. 63--84, Apr. 2014.

\bibitem{ambi16}
S.~Ambikasaran, D.~Foreman-Mackey, L.~Greengard, D.~W. Hogg, and M.~O'Neil,
  ``Fast direct methods for gaussian processes,'' \emph{IEEE Transactions on
  Pattern Analysis \& Machine Intelligence}, vol.~38, no.~2, pp. 252--265,
  2016.

\bibitem{borm05interpolation}
S.~B{\"o}rm, M.~L{\"o}hndorf, and J.~M. Melenk, ``Approximation of integral
  operators by variable-order interpolation,'' \emph{Numerische Mathematik},
  vol.~99, no.~4, pp. 605--643, 2005.

\bibitem{hackbush_2000}
W.~Hackbusch and B.~N. Khoromskij, ``\BIBforeignlanguage{English}{A sparse
  {$\mathcal{H}$}-matrix arithmetic. {P}art {II}: Application to
  multi-dimensional problems},''
  \emph{\BIBforeignlanguage{English}{Computing}}, vol.~64, no.~1, pp. 21--47,
  2000.

\bibitem{hackbush_2003}
L.~Grasedyck and W.~Hackbusch, ``Construction and arithmetics of
  {$\mathcal{H}$}-matrices,'' \emph{Computing}, vol.~70, no.~4, pp. 295--334,
  2003.

\bibitem{abdelfattah16a}
A.~Abdelfattah, D.~Keyes, and H.~Ltaief, ``Kblas: An optimized library for
  dense matrix-vector multiplication on gpu accelerators,'' \emph{ACM
  Transactions Mathematical Software}, vol.~42, no.~3, pp. 18:1--18:31, 2016.

\bibitem{abdelfattah16b}
A.~Abdelfattah, H.~Ltaief, D.~Keyes, and J.~Dongarra, ``Performance
  optimization of sparse matrix-vector multiplication for multi-component
  pde-based applications using gpus,'' \emph{Concurrency and Computation:
  Practice and Experience}, vol.~28, no.~12, pp. 3447--3465, 2016.

\bibitem{cusparse16}
NVIDIA, \emph{{cuSPARSE} Library Documentation (v8.0)}, 2016.

\bibitem{borm_2010}
S.~B{\"o}rm, \emph{Efficient numerical methods for non-local operators:
  {$\mathcal{H}^2$}-matrix compression, algorithms and analysis}, ser. EMS
  tracts in mathematics.\hskip 1em plus 0.5em minus 0.4em\relax European
  Mathematical Society, 2010, vol.~14.

\end{thebibliography}



\end{document}